\documentclass[11pt]{article}

\usepackage[final]{acl}
\usepackage{hyperref}

\usepackage{times}
\usepackage{latexsym}

\usepackage[T1]{fontenc}

\usepackage[utf8]{inputenc}

\usepackage{microtype}

\usepackage{inconsolata}

\usepackage{graphicx}
\usepackage[table]{xcolor}

\usepackage{amssymb} 
\usepackage{amsfonts}
\usepackage{amsmath, amsthm} 

\usepackage{enumitem}
\usepackage{xcolor}
\usepackage{tabularx}
\usepackage{colortbl}
\usepackage{orcidlink}

\definecolor{midgreen}{RGB}{0,150,0}
\definecolor{midred}{RGB}{250,0,0}
\definecolor{lime}{rgb}{0.88,2,10}

\usepackage[table]{xcolor}

\usepackage[labelformat=simple]{subcaption}

\newcommand{\fref}[1]{Fig.~\ref{#1}}

\newcommand{\sref}[1]{Section~\ref{#1}}

\newcommand*{\Resize}[2]{\resizebox{#1}{!}{$#2$}}%

\usepackage{booktabs}
\usepackage{multirow}
\usepackage{multicol}
\usepackage{tabularx}

\usepackage[linesnumbered,ruled,vlined]{algorithm2e}

\usepackage{eso-pic}
\usepackage{xcolor}
\title{SecureGate: Learning When to Reveal  PII Safely via Token-Gated Dual-Adapters for Federated LLMs}

\author{
Mohamed Shaaban \and Mohamed Elmahallawy\thanks{Corresponding author.}\\
Washington State University \\
\texttt{\{mohamed.shaaban, mohamed.elmahallawy\}@wsu.edu}
}

\begin{document}
\maketitle
\AddToShipoutPictureFG*{%
  \AtPageLowerLeft{%
 
    \hspace*{0.08\paperwidth}%
    \raisebox{1.7cm}[0pt][0pt]{%
      \parbox{0.84\paperwidth}{\footnotesize
      \noindent  \center {\em To appear in the Main Proceedings of ACL 2026:
      The 64th Annual Meeting of the Association for Computational Linguistics.}}
    }%
  }%
}
\vspace{0.5em}
 
\begin{abstract}

Federated learning (FL) enables collaborative training across organizational silos without sharing raw data, making it attractive for privacy-sensitive applications. With the rapid adoption of large language models (LLMs), federated fine-tuning of generative LLMs has gained attention as a way to leverage distributed data while preserving confidentiality. However, this setting introduces fundamental challenges: (i) \emph{privacy leakage} of personally identifiable information (PII) due to LLM memorization, and (ii) a persistent \emph{tension between global generalization and local utility} under heterogeneous data. Existing defenses, such as data sanitization and differential privacy, reduce leakage but often degrade downstream performance. We propose \textsc{SecureGate}, a \emph{privacy-aware federated fine-tuning framework} for LLMs that provides fine-grained privacy control without sacrificing utility. \textsc{SecureGate} employs a dual-adapter LoRA architecture: a \emph{secure adapter} that learns sanitized, globally shareable representations, and a \emph{revealing adapter} that captures sensitive, organization-specific knowledge. A token-controlled \emph{gating module} selectively activates these adapters at inference time, enabling controlled information disclosure without retraining. Extensive experiments across multiple LLMs and real-world datasets show that \textsc{SecureGate} improves task utility while substantially reducing PII leakage, achieving up to a 31.66$\times$ reduction in inference attack accuracy and a 17.07$\times$ reduction in extraction recall for {\em unauthorized} requests. Additionally, it maintains 100\% routing reliability to the correct adapter and incurs only minimal computational and communication overhead. Code is available at \url{https://github.com/wsu-cyber-security-lab-ai/SecureGate}.





\end{abstract}

\section{Introduction}

In the era of large language models (LLMs), organizations across industry and academia increasingly rely on pre-trained models to support domain-specific applications such as scientific analysis, healthcare, and enterprise workflows~\cite{ling2023domain, lu2025fine}. Training LLMs from scratch is often prohibitively expensive in terms of computation, data, and engineering effort; consequently, fine-tuning pre-trained models has become the dominant approach for domain and task adaptation~\cite{2025_Pratap}. While effective, independently fine-tuned models remain confined to local data silos, preventing organizations from benefiting from complementary knowledge distributed across institutions and often resulting in limited generalization and robustness~\cite{lin-etal-2022-fednlp}.

Federated learning (FL)~\cite{kairouz2021advances} provides a principled framework to address this limitation by enabling collaborative model training without sharing raw data. In a standard FL setup, each participant fine-tunes a local copy of a shared pre-trained model using private data and periodically transmits model updates to a central aggregator for global aggregation. Although FL has demonstrated success for conventional machine learning and smaller neural models, extending it to large-scale generative LLMs poses fundamental challenges. LLMs are prone to {\em memorizing sensitive information}~\cite{carlini2021extracting}, and naive aggregation of model updates can {\em leak personally identifiable information (PII)}~\cite{ijcai2025p1156}. Moreover, aggregating updates often favors average global performance, degrading utility on participants’ domain-specific data and intensifying the tension between global generalization and local personalization~\cite{parthasarathy2024ultimate}. 

To reduce PII leakage during collaborative fine-tuning, organizations commonly rely on defenses such as data scrubbing~\cite{akkus2025generated} and differential privacy (DP)~\cite{hassan2019differential}. While effective at limiting memorization-based leakage, these methods often degrade model utility by removing task-relevant context or injecting excessive noise, particularly as privacy constraints tighten. In practice, this trade-off can negate the benefits of cross-organization collaboration, yielding global models that underperform locally trained alternatives. Moreover, existing defenses scale poorly to large generative LLMs and rarely address personalization, communication efficiency, and client heterogeneity jointly. Thus, this work addresses the following question: \emph{How can federated fine-tuning of generative LLMs simultaneously ensure strong privacy guarantees, high global and local utility, and practical efficiency?}

We propose \textsc{SecureGate}, a federated fine-tuning framework for organizations adapting LLMs in privacy-sensitive settings. \textsc{SecureGate} employs two lightweight (Low Rank Adapters) LoRAs~\cite{hu2022lora}: a \emph{secure adapter} that learns sanitized, shareable representations, and a \emph{revealing adapter} that retains sensitive, organization-specific knowledge. A token-controlled gating module selectively routes prompts at inference time, enabling fine-grained access control without retraining. By aligning global and local representations, \textsc{SecureGate} enables personalized adaptation while maintaining communication efficiency and scalability. We summarize our contribution as:
\vspace{-5mm}

\begin{enumerate}[leftmargin=*,label={\bf \arabic*.}, leftmargin=0pt, labelwidth=*, labelsep=0.5em, align=left]
\item \textbf{Token-controlled privacy paradigm (\textsc{SecureGate}).}  
We introduce a token-based FL paradigm that protects private organizational knowledge by \emph{locking sensitive personalized components}. Access to these components is granted only when a learned \fcolorbox{green!50!black}{green!15}{\raisebox{-0.2ex}{\textcolor{midgreen}{\texttt{[Special-Token]}}}} is presented; unauthorized parties—including the aggregator and other clients—cannot access or infer the protected information.

\item \textbf{Dual-adapter  access-controlled architecture.}  
\textsc{SecureGate} realizes model adaptation through: 
    \begin{itemize}[leftmargin=*]
    \item \emph{Secure adapter:} Encodes globally shareable knowledge while preserving PII using privacy defenses such as noise injection (e.g., DP) or data scrubbing (e.g., adding \fcolorbox{red!50}{pink!20}{\raisebox{-0.2ex}{\textcolor{midred}{\texttt{[MASK]}}}}).

   \item \emph{Revealing adapter:} Encodes highly sensitive, organization-specific knowledge and is activated only for queries from authorized organizations that include a valid \fcolorbox{green!50!black}{green!15}{\raisebox{-0.2ex}{\textcolor{midgreen}{\texttt{[Special-Token]}}}}.

    \end{itemize}

  \item \textbf{Privacy-aware gating and routing.}  A lightweight gating module dynamically routes prompts to the appropriate adapter based on token signals, enabling selective disclosure at inference time: authorized queries access sensitive knowledge, while unauthorized queries rely solely on the secure adapter.

   \item \textbf{Empirical evaluation.} We evaluate \textsc{SecureGate} on multiple LLMs and real-world datasets (e.g., ECHR, Yelp Reviews). Results show strong privacy--utility trade-offs: {\em authorized} performance reaches 25.20\% inference accuracy and 6.32 perplexity (PPL) score, while {\em unauthorized} access is suppressed to a leakage floor of 4.20\% and 15.89 PPL. Overall, \textsc{SecureGate} achieves up to a \(31.66\times\) reduction in inference accuracy and a \(17.07\times\) reduction in extraction recall for {\em unauthorized} requests, while preserving low computation \&communication cost and \(100\%\) routing reliability.

  
\end{enumerate}

\vspace{-4mm}

\section{Related Work}
\vspace{-1mm}

\subsection{Federated Learning for LLMs}\label{sec:FL_LLMs}
FL enables collaborative training across distributed clients without sharing raw data. Foundational methods such as FedAvg~\cite{mcmahan2017communication}, FedAdagrad, FedAdam, FedYogi~\cite{reddi2020adaptive}, and FedAvgM~\cite{hsu2019measuring} provide techniques for aggregating client updates and handling heterogeneous data. Early work addressed statistical heterogeneity, while recent approaches like HPFL~\cite{shenhot} and LG-Mix~\cite{jiang2024heterogeneous} decompose models into shared backbones and personalized components. 

The rise of LLMs has motivated adapting classical FL approaches to manage massive parameters and memory across clients while enabling collaborative training of globally shared LLMs. Directly transmitting full LLM weights is infeasible due to prohibitive communication and storage costs. Lightweight adaptation techniques such as LoRA mitigate this challenge: FedLoRA~\cite{wu2024fedlora} decomposes each layer into a shared full-rank component and a client-specific low-rank adapter to reduce non-IID effects, while pFedLoRA~\cite{yi2023pfedlora} iteratively trains local adapters across heterogeneous clients' architectures, aggregating them into a global adapter with minimal communication. These methods make federated LLM training practical, but they primarily optimize utility, {\em leaving models vulnerable to memorization and unintended disclosure of sensitive local data.}

\subsection{Privacy Leakage in Federated LLMs}

Federated LLMs are still in their infancy, but initial efforts have begun to address privacy concerns. Personalized FL (PFL) methods aim to improve generalization while limiting information leakage by combining shared knowledge with client-specific adaptations. For example, FedDPA~\cite{long2024dual} introduces global–local adapter pairs dynamically fused at inference, yet it does not provide inherent access control for sensitive data.

On the security side, cryptographic techniques such as secure aggregation have been extensively explored in classical FL~\cite{allouah2025towards, zuo2024pack, pu2025janus, mansouri2023sok}, but their application to federated LLMs remains limited. Most focus solely on training-time protection, leaving inference-time privacy largely unaddressed. Moreover, these methods incur substantial computation and communication overheads, which would scale poorly to LLM-sized models.  FDLoRA~\cite{qi2024fdlora} takes an early step by separating global and local adapters, but it provides no privacy guarantees during inference. 

Building on these insights, we propose \textsc{SecureGate}, a framework that combines adapter-based personalization with token-controlled, per-request selective disclosure, {\em enabling fine-grained, secure access to federated LLM knowledge while preserving both training- and inference-time privacy and maintaining high local and global model utility.}

\section{Problem Formulation \& Threat Model}
\subsection{Problem Formulation}
\noindent\textbf{Personalized Federated Learning.}  
FL enables $N$ organizations, each with a private dataset $D_n$ ($n=1,\dots,N$), to collaboratively train a shared model without exchanging raw data. At each communication round $t$, organizations perform local training starting from the current global parameters $\boldsymbol{w}^t$, producing updated local parameters $\boldsymbol{w}_n^{t+1}$. The server aggregates these updates (e.g., via FedAvg~\cite{mcmahan2017communication}) to obtain the next global model:
$\boldsymbol{w}^{t+1} = \sum_{n=1}^{N} \frac{|D_n|}{|\mathcal{D}|} \boldsymbol{w}_n^{t+1}$, and broadcasts it to all organizations until convergence.

Personalized FL (PFL)~\cite{liu2024recent} extends FL to preserve per-organization utility by maintaining both shared and local parameters: $\boldsymbol{w}_n^t = \boldsymbol{w}^t + \boldsymbol{w}_{p,n}^t$, where $\boldsymbol{w}_{p,n}^t$ is the personalized component. A typical objective is
\begin{equation}
\min_{\boldsymbol{w}^t,\{\boldsymbol{w}_{p,n}^t\}} \sum_{n=1}^N \Big(\mathcal{L}_n(\boldsymbol{w}^t + \boldsymbol{w}_{p,n}^t) + \lambda\,\Omega(\boldsymbol{w}_{p,n}^t)\Big),
\end{equation}
with $\mathcal{L}_n$ the local loss, $\Omega$ a regularizer, and $\lambda$ a trade-off weight. Local updates follow standard gradient steps:
$\boldsymbol{w}_{p,n}^{t+1} = \boldsymbol{w}_{p,n}^t - \eta \nabla \mathcal{L}_n(\boldsymbol{w}^t + \boldsymbol{w}_{p,n}^t)$.

PFL can be realized via mixture-of-experts, meta-learning, or parameter-efficient fine-tuning, with personalized parameters typically retained locally to limit exposure. This inherent tension between global generalization and per-organization specialization motivates our work to jointly balance privacy and personalization while {\em maximizing both global and local model utility.}

\noindent\textbf{Parameter-Efficient Adaptation of LLMs.}  Training an LLM from scratch involves billions of parameters and is often unnecessary when adapting to a specific domain. Full fine-tuning can also drift from pretrained knowledge and is computationally expensive. Parameter-efficient fine-tuning (PEFT) methods address this by updating only small, task-specific components while keeping the base model frozen. A common approach is LoRA, which trains compact adapter matrices instead of all model weights, reducing computation, storage, and communication costs. Formally, for a pretrained weight matrix $\boldsymbol{w}_0 \in \mathbb{R}^{d \times k}$, LoRA decomposes the weight update as $\boldsymbol{w} = \boldsymbol{w}_0 + \Delta \boldsymbol{w}, ~
\Delta \boldsymbol{w} = \boldsymbol{B}\boldsymbol{A}, \; \boldsymbol{A} \in \mathbb{R}^{r \times k}, \; \boldsymbol{B} \in \mathbb{R}^{d \times r}, \; r \ll \min(d,k)$,
where only $\boldsymbol{A}$ and $\boldsymbol{B}$ are trainable. The forward pass becomes $h = \boldsymbol{w} x = \boldsymbol{w}_0 x + \boldsymbol{B}\boldsymbol{A}x$, preserving the pretrained base while adapting to task-specific knowledge. Distinct LoRA adapters can be learned for different tasks and selectively combined with the frozen model at inference, enabling efficient, modular personalization.

\noindent\textbf{When LLMs meet FL.}
 Leveraging advances in FL, PFL, and LoRA adapters, federated personalized fine-tuning lets organizations adapt pretrained LLMs to domain data without sharing raw corpora. In the PFL-for-LLMs setting each organization $n$ keeps a global model $\boldsymbol{w}$ and a compact personalized adapter $\Delta \boldsymbol{w}_{p,n}$, often parameterized with low-rank (LoRA) updates:
$\Delta \boldsymbol{w}_{p,n}\approx\sum_{\ell}\mathbf{B}_{n,\ell}\mathbf{A}_{n,\ell}$, where ${\ell}$ is an index that iterates over the layers (or parameterized components). Thus, communicating only the adapter parameters greatly reduces storage and bandwidth. However, directly applying FL/PFL to LLMs is problematic: {\em large models tend to memorize sensitive data (risking PII leakage), their checkpoints and updates are communication-intensive, and organization heterogeneity and regulatory constraints undermine naive FL/PFL averaging}. These limitations motivate \textsc{SecureGate}, designed to curb leakage while preserving global and per-organization utility.

\subsection{Threat Model}

We consider realistic adversaries in federated LLM deployments, where organizations retain raw data locally and share only model updates or adapter weights. Adversaries aim to extract sensitive information or degrade global/local model utility.

\noindent \textbf{Roles, Capabilities, and Goals.} We consider three principal adversary types: 
(i) \emph{honest-but-curious server} — follows the protocol but inspects updates or checkpoints;  (ii) \emph{curious collaborating organizations} — analyze exchanged parameters to extract proprietary or PII information;  (iii) \emph{external adversaries} — eavesdrop on communications to perform membership inference. Adversaries may inspect updates or checkpoints, query models via APIs, and combine auxiliary public or leaked data with observations. Their goals are to reconstruct or extract private data (memorization/extraction attacks) and perform membership or attribute inference on organizational datasets.

We also assume adversaries do \emph{not} have direct access to raw datasets but may observe model artifacts and query outputs.

\begin{figure*}[!t]
  \centering
  \includegraphics[width=0.95\textwidth]{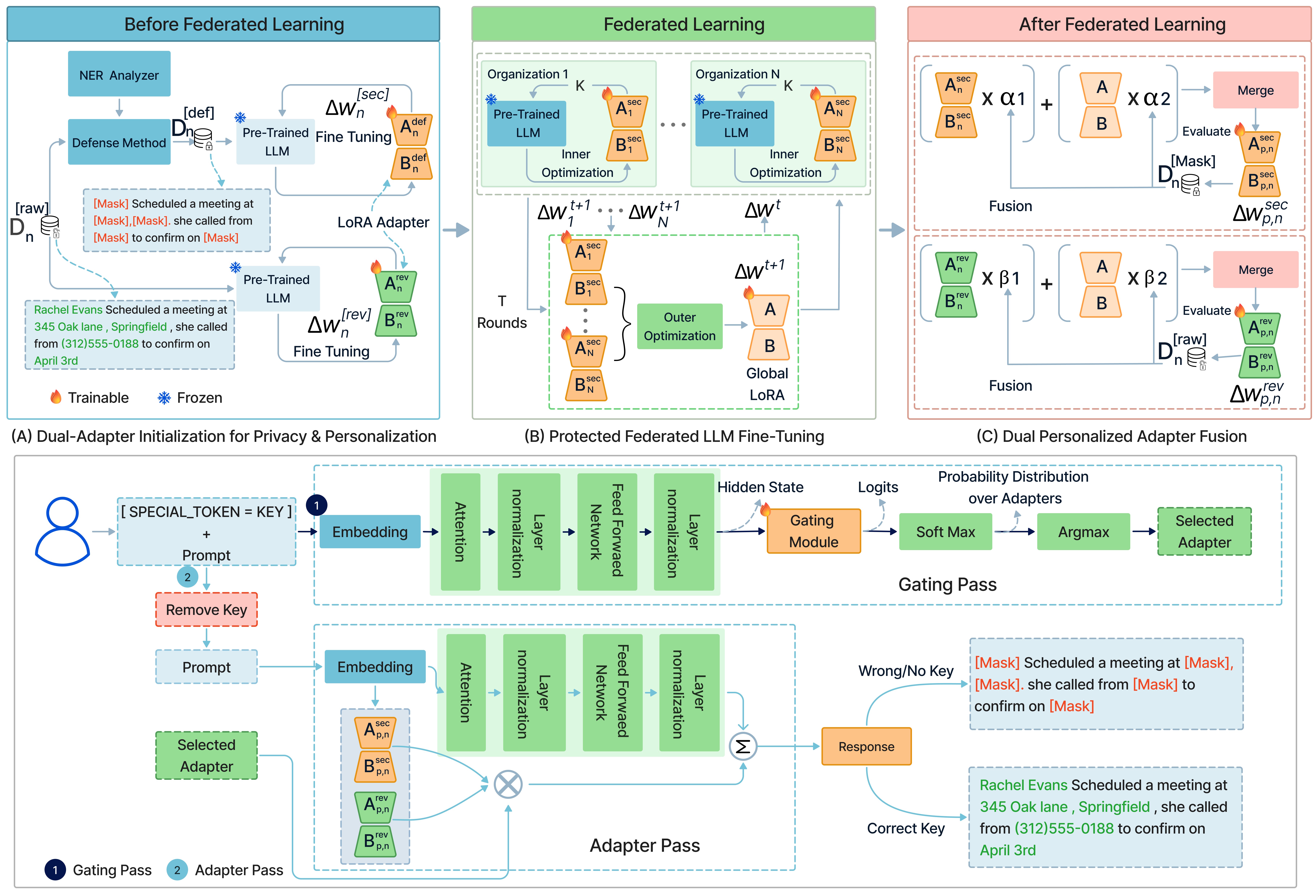}
  \caption{An Illustration of the \textsc{SecureGate} Framework.}\vspace{-4.5mm}
  \label{fig:methodology}
\end{figure*}

\section{Methodology}
This section introduces \textsc{SecureGate}, a framework that addresses privacy leakage in federated LLMs through three components: {\em a learnable authorization token, a dual-adapter LoRA design, and a gating module for selective adapter activation}. 
The framework operates as follows (see~\fref{fig:methodology}):






\subsection{Dual-Adapter Initialization for Privacy \& Personalization}


In this stage, \textsc{SecureGate} allows each organization $n \in \mathcal{N}$ to initialize and fine-tune dual LoRA adapters: (i) a secure adapter that generates representations sanitized of PII, and (ii) a revealing adapter for controlled disclosure (see \fref{fig:methodology}.a). Let $\mathcal{D}_n$ denotes organization $n$'s local dataset, which is partitioned into two views: a \textit{raw view} $\mathcal{D}_n^{\text{raw}}$ and a \textit{masked view} $\mathcal{D}_n^{\text{mask}}$, where PII is automatically detected (e.g., via named-entity recognition, NER~\cite{nadeau2007survey}) and replaced with \fcolorbox{red!50}{pink!20}{\raisebox{-0.2ex}{\textcolor{midred}{\texttt{[MASK]}}}} using a defense scheme (e.g., scrubbing~\cite{pilan2022text}). We define two parameter-efficient LoRA updates per organization:
\begin{equation}
\label{eq:lora_initalize}
\Resize{6.9cm}{\Delta \boldsymbol{w}_{n}^{(\text{rev})} = \boldsymbol{B}_{n}^{(\text{rev})} \boldsymbol{A}_{n}^{(\text{rev})},~~
\Delta \boldsymbol{w}_{n}^{(\text{sec})} = \boldsymbol{B}_{n}^{(\text{sec})} \boldsymbol{A}_{n}^{(\text{sec})},}
\end{equation}
where $\Resize{1cm}{\Delta \boldsymbol{w}_{n}^{(\text{rev})}}$ and $\Resize{1cm}{\Delta \boldsymbol{w}_{n}^{(\text{sec})}}$ are the revealing and secure adapters trained on $\Resize{0.66cm}{\mathcal{D}_n^{\text{raw}}}$ and $\Resize{0.67cm}{\mathcal{D}_n^{\text{mask}}}$, respectively.

During fine-tuning, only $\Resize{1cm}{\Delta \boldsymbol{w}_{n}^{(\text{sec})}}$ is shared and aggregated by the server, ensuring the global model encodes only sanitized knowledge. In contrast, $\Resize{1cm}{\Delta \boldsymbol{w}_{n}^{(\text{rev})}}$ remains local and, after fusion with the final global model, is accessed exclusively via token-gated control (\sref{sec:gating}), allowing only authorized users to access sensitive information.



\subsection{Protected Federated LLM Fine-Tuning}\label{sec:Fl_loop}

We employ a two-level optimization scheme that alternates between a server-side \emph{outer loop} and an organization-side \emph{inner loop} (\fref{fig:methodology}.b). The goal is to learn a global LoRA adapter that captures domain-generalizable knowledge while allowing each organization to maintain a personalized variant for local utility, {\em ensuring that no organization’s PII is exposed to others or the server.}  

At communication round $t\in\{1,\dots,T\}$, the server broadcasts the current global secure adapter $\Delta \boldsymbol{w}^{t}$. Each organization $n\in\mathcal{N}$ then performs $K$ inner-loop updates on its masked dataset $\mathcal{D}_n^{\text{mask}}$ using AdamW~\cite{loshchilov2017decoupled}, producing locally updated adapters $\Delta \boldsymbol{w}_n^{t+1}$, which are sent to the server. Next, the server aggregates these updates via weighted averaging with Nesterov momentum~\cite{sutskever2013importance}:\vspace{-1mm}
\begin{equation}
\Delta \bar{\boldsymbol{w}}^{t+1} =
\sum_{n=1}^{N}
\frac{|\mathcal{D}_n^{\text{mask}}|}{\sum_{k=1}^{N} |\mathcal{D}_k^{\text{mask}}|}
\Delta \boldsymbol{w}_n^{t+1},
\label{eq:outer_loop}\vspace{-2mm} 
\end{equation}
This average is combined with a scaled momentum vector \(\Delta \boldsymbol{v}^t\) to form a lookahead position as $\Delta \boldsymbol{p}^t = \Delta \boldsymbol{w}^t + m \Delta \boldsymbol{v}^t$, where \(m\) controls the momentum strength. The momentum is then updated by moving towards the difference between the weighted average and lookahead scaled by the learning rate \(\eta\) as:\vspace{-1mm}
\begin{equation} \label{eq:momentem_based} \Delta \boldsymbol{v}^{t+1} = m \Delta \boldsymbol{v}^t + \eta \left( \Delta \bar{\boldsymbol{w}}^{t+1} - \Delta \boldsymbol{p}^t \right)\vspace{-2mm} 
\end{equation}
Finally, the server updates the global model with the new momentum: \vspace{-1mm}
\begin{equation} \label{eq:global_update_using_momentum} \Delta\boldsymbol{w}^{t+1} = \Delta\boldsymbol{w}^t + \Delta \boldsymbol{v}^{t+1}\vspace{-2mm}
\end{equation}
By restricting communication to lightweight secure adapters, this scheme significantly reduces communication overhead while preserving strong global generalization and enabling privacy-preserving local adaptation at each organization.

\vspace{-1mm}
\subsection{Dual Personalized Adapter Fusion}\label{sec:fusion}

After completing federated fine-tuning (\sref{sec:Fl_loop}), each organization fuses the final global adapter $\Delta \boldsymbol{w}^{T}$ with its local secure and revealing adapters. This yields two organization-specific adapters: (i) a \emph{secure personalized adapter} that preserves privacy while maintaining global utility, and (ii) a \emph{revealing personalized adapter} that selectively discloses sensitive local knowledge to authorized users (see \fref{fig:methodology}.c).

\noindent\textbf{Secure Personalized Adapter.}
Let $\Delta\boldsymbol{w}_{n}^{T}$  denotes the locally trained secure adapter of organization $n$ after $T$ rounds. The secure personalized adapter is obtained via linear fusion:\vspace{-1.5mm}
\begin{equation}\vspace{-2.5mm}
\Delta\boldsymbol{w}^{(\text{sec})}_{p,n}
= \alpha_{1}\Delta\boldsymbol{w}^{T}
+ \alpha_{2}\Delta\boldsymbol{w}_{n}^{T},
\label{eq:secure_fusion}
\end{equation}
where $\alpha_1$ and $\alpha_2$ balance global generalization and local adaptation. Following~\cite{huang2023lorahub}, the fusion weights are learned via black-box optimization by minimizing the cross-entropy loss on a small query set $Q\subset\mathcal{D}_n^{\text{mask}}$ with $\ell_1$ regularization:\vspace{-1.5mm}
\begin{equation}\vspace{-2mm}
\min_{\{\alpha_i\}}
\;\mathcal{L}(\alpha; Q)
+ \psi \sum_i |\alpha_i|.
\label{eq:fusion_obj}
\end{equation}
where $\psi$ regulates the strength of regularization. Since $\Delta\boldsymbol{w}_{n}^{T}$ is trained on masked data (e.g., PII replaced with \fcolorbox{red!50}{pink!20}{\raisebox{-0.2ex}{\textcolor{midred}{\texttt{[MASK]}}}}), this fusion preserves privacy while integrating global knowledge.

\vspace{1mm}
\noindent\textbf{Revealing Personalized Adapter.}
In parallel, each organization fuses $\Delta\boldsymbol{w}^{T}$ with its local revealing adapter $\Delta\boldsymbol{w}_{n}^{(\text{rev})}$, trained on raw data:\vspace{-1.5mm}
\begin{equation}\vspace{-2.5mm}
\Delta\boldsymbol{w}^{(\text{rev})}_{p,n}
= \beta_{1}\Delta\boldsymbol{w}^{T}
+ \beta_{2}\Delta\boldsymbol{w}_{n}^{(\text{rev})},
\label{eq:reveal_fusion}
\end{equation}
where $\beta_1$ and $\beta_2$ are optimized similar to  $\alpha_i$ in Eq.~\eqref{eq:fusion_obj}. This adapter remains strictly local and is accessed only through token-controlled gating at inference time, enabling controlled disclosure of sensitive organization-specific knowledge.

Together, these adapters allow organizations to benefit from shared federated knowledge while preserving privacy and enabling fine-grained, authorization-based information access.

\vspace{-1mm}

\subsection{Gating Module Training}\label{sec:gating}
\vspace{-1mm}
To enable controlled disclosure at inference time, we introduce a lightweight \emph{gating module} that routes each query to either the \emph{secure} or \emph{revealing} personalized adapter (\sref{sec:fusion}). Together with the dual adapters, the gating module forms the access-control core of \textsc{SecureGate}, enabling per-request, key-controlled disclosure while preserving shared model utility (see \fref{fig:methodology}.d).

\noindent\textbf{Design and Input Encoding.}
The gating module is a small MLP designed for minimal local overhead. Each query is prefixed with a learnable access key encoded as a special token: \fcolorbox{green!50!black}{green!15}{\raisebox{-0.2ex}{\textcolor{midgreen}{\texttt{[SPECIAL\_TOKEN=KEY]}\;\texttt{PROMPT}}}}. This key is added as an atomic token to an organization-local tokenizer, while a public tokenizer excludes it, preventing token splitting and leakage (see  Appendix~\ref{subsec:key-security}). The gating module operates on the final-layer hidden state $h_{\text{key}}$ of the key token extracted from the frozen base model, which we find provides the most robust routing signal (see Appendix~\ref{subsec:gating_feature_ablation}).

\noindent The gating logits are computed as
\begin{equation}\vspace{-1.5mm}
z_i = \mathrm{MLP}(h_{\text{key}}), \quad i \in \mathcal{I},
\label{eq:hidden_extraction}
\end{equation}
followed by a softmax:
$p_i = \frac{\exp(z_i)}{\sum_{j\in\mathcal{I}} \exp(z_j)}$.

The adapter is selected as $a^*=\arg\max_i p_i$, with an optional threshold $\tau$ to default to the secure adapter when confidence is low. Notably, this design naturally extends to multiple adapters with distinct keys and access levels.


\vspace{1mm}
\noindent\textbf{Gating Module Training.} Each organization trains its gating module locally using a cross-entropy loss: $ \mathcal{L}_{\text{gating}} = -\sum_{i} y_i \log p_i$, where $y_i$ indicates the correct adapter for a given key. Training data is {\em synthetically} generated to cover authorized and unauthorized cases (details on synthetic data generation are provided in Appendix~\ref{subsec:synthetic_routing_data}). During training, the base model and adapters remain frozen, confining learning to the small gating network. The trained gating module remains local and can be efficiently retrained when keys are rotated, {\em without modifying the model or adapters.}

\vspace{1mm}
\noindent\textbf{Two-Pass Inference.}
To prevent key contamination of outputs, \textsc{SecureGate} uses a two-pass inference procedure. In the \emph{gating pass}, the full prompt (with key) is processed by the frozen base model to extract $h_{\text{key}}$ and select an adapter. In the \emph{adapter pass}, the key is removed and the cleaned prompt is re-evaluated using the selected adapter to generate the final response. This ensures reliable routing without key-induced bias in generation.

\section{Performance Evaluation}\label{sec:rlt}
\subsection{Experimental Setup}

\noindent\textbf{Datasets.} We evaluate \textsc{SecureGate} on two benchmark datasets: (i) \textit{Yelp Reviews~\cite{zhang2015character}}, a large collection of user-generated reviews with text, star ratings, and metadata such as timestamps and user/business IDs; and (ii) \textit{ECHR~\cite{chalkidis2019neural}}, comprising legal case documents from the European Court of Human Rights containing personal information. We uniformly sample 10,000 instances from each dataset to create balanced subsets. Our FL setup includes 10, 20, or 30 organizations, each training on a distinct 10\% partition of the sampled data.


\noindent\textbf{LLM Models.} We experiment with SOTA open-source LLMs Qwen/Qwen~3-1.7B~\cite{qwen3_2025}, Google's Gemma~2-2b~\cite{gemma_2024} and Meta’s Llama~3.2-1B/3B~\cite{dubey2024llama}, fine-tuned via LoRA adapters.

\noindent\textbf{Hyperparameters.} For LoRA training, we set the rank \( r \in \{4, 8, 12, 16\} \). Each client fine-tunes its local parameters for three epochs using AdamW with learning rate $\eta_{\text{local}}=10^{-4}$. For global aggregation, we use $\eta_{\text{global}} = 0.01$ and \(T = 20\) rounds. Remaining parameters and implementation specifics are detailed in Appendix~\ref{sec:experimental_details}.

\noindent\textbf{Attack Types.} We evaluate privacy leakage under two threats: (i) \textit{Extraction Attack:} The model is sampled to generate text, after which NER is applied to extract PII; leakage is measured by comparing extracted entities against a baseline model; (ii) \textit{Inference Attack:} Given a candidate set of $c$ PII values ($c{=}50$ unless stated otherwise), the attacker selects the value that minimizes model perplexity for a fixed prefix–suffix context, assuming access to an auxiliary set of 100 PII-containing statements.

\noindent{\bf Evaluation Metrics.} We evaluate model performance and privacy leakage using standard metrics, including Precision, Recall, Perplexity, and computational efficiency (FLOPs); formal definitions are provided in Appendix \ref{subsec:metric_definitions}.

\begin{table*}[t!]
\centering
\caption{Client-wise comparative analysis of inference attack accuracy (\%): evaluating \textsc{SecureGate} in authorized vs.\ unauthorized scenarios across diverse FL environments on the ECHR dataset using the Llama-1B model.} 
\vspace{-2mm}
\label{tab:fl_full_comparison_no_rotate_main}
\small 
\setlength{\tabcolsep}{4.5pt} 
\renewcommand{\arraystretch}{1.1}
\definecolor{highlight}{gray}{0.93}
\definecolor{securegateblue}{rgb}{0.9, 0.95, 1.0}
\resizebox{\linewidth}{!}{
\begin{tabular}{@{} l cccccc @{\hskip 0.3in} cccccc @{}}
\toprule
&  \multicolumn{6}{c}{\textbf{Authorized Access} (Correct Token) {\bf $\uparrow$}} & \multicolumn{6}{c}{\textbf{Unauthorized Access} (Wrong/No Token) {\bf $\downarrow$}} \\
\cmidrule(r{0.3in}){2-7} \cmidrule{8-13}
\textbf{Client} & \textbf{FedAdagrad} & \textbf{FedAvg} & \textbf{FedAdam} & \textbf{FedYogi} & \textbf{FedAvgM} & \cellcolor{securegateblue}\textbf{SecureGate} 
& \textbf{FedAdagrad} & \textbf{FedAvg} & \textbf{FedAdam} & \textbf{FedYogi} & \textbf{FedAvgM} & \cellcolor{securegateblue}\textbf{SecureGate} \\
\midrule
1  & 14.10 & 20.00 & 26.74 & 30.12 & 24.14 & \cellcolor{securegateblue}\textbf{35.96} & 4.82 & 4.60 & 8.60 & 6.67 & 3.16 & \cellcolor{securegateblue}3.61 \\
2  & 15.00 & 14.10 & 22.08 & 22.50 & 20.48 & \cellcolor{securegateblue}\textbf{25.64} & 4.65 & 5.81 & 5.75 & 4.88 & 5.75 & \cellcolor{securegateblue}5.56 \\
3  & 9.76  & 11.76 & 20.24 & 21.59 & 20.45 & \cellcolor{securegateblue}\textbf{26.74} & 2.33 & 1.15 & 8.14 & 1.15 & 1.12 & \cellcolor{securegateblue}4.65 \\
4  & 11.39 & 16.46 & 17.65 & 21.25 & 18.82 & \cellcolor{securegateblue}\textbf{21.79} & 3.75 & 6.82 & 5.75 & 3.49 & 2.17 & \cellcolor{securegateblue}4.94 \\
5  & 20.25 & 21.43 & 16.47 & 13.10 & 19.10 & \cellcolor{securegateblue}\textbf{29.87} & 7.23 & 4.55 & 6.32 & 2.20 & 2.11 & \cellcolor{securegateblue}2.22 \\
6  & 16.46 & 19.32 & 19.54 & 19.51 & 18.29 & \cellcolor{securegateblue}\textbf{25.00} & 2.67 & 6.98 & 4.35 & 3.45 & 6.74 & \cellcolor{securegateblue}3.80 \\
7  & 7.41  & 13.19 & 10.71 & 18.60 & 15.56 & \cellcolor{securegateblue}\textbf{28.05} & 4.71 & 7.45 & 3.45 & 2.13 & 3.09 & \cellcolor{securegateblue}5.75 \\
8  & 11.76 & 17.86 & 21.33 & 12.50 & 15.12 & \cellcolor{securegateblue}\textbf{29.87} & 6.25 & 4.55 & 4.55 & 1.11 & 1.05 & \cellcolor{securegateblue}7.79 \\
9  & 11.39 & 16.85 & 22.22 & 20.93 & 20.69 & \cellcolor{securegateblue}\textbf{32.10} & 6.33 & 5.38 & 5.21 & 4.60 & 7.69 & \cellcolor{securegateblue}1.22 \\
10 & 16.46 & 18.99 & 14.81 & 20.00 & 28.57 & \cellcolor{securegateblue}\textbf{35.90} & 2.44 & 4.65 & 5.56 & 7.61 & 6.25 & \cellcolor{securegateblue}2.44 \\
\midrule
\rowcolor{highlight}
\textbf{Average} & \textbf{13.40} & \textbf{17.00} & \textbf{19.18} & \textbf{20.01} & \textbf{20.12} & \cellcolor{securegateblue}\textbf{25.20} & \textbf{4.52} & \textbf{5.19} & \textbf{5.77} & \textbf{3.73} & \textbf{3.92} & \cellcolor{securegateblue}\textbf{4.20} \\
\bottomrule
\end{tabular}
}
\end{table*}

\subsection{Experimental Results}

\begin{table*}[!t]
\centering
\caption{\resizebox{5.8in}{!}{\textsc{SecureGate} 
robustness: inference and extraction attacks across models, datasets, and access scenarios.}}

\vspace{-2mm}
\label{tab:combined_results}
\setlength{\tabcolsep}{1.5pt}  
\renewcommand{\arraystretch}{1}
\definecolor{secureblue}{rgb}{0.9, 0.95, 1.0}
\definecolor{securered}{rgb}{1.0, 0.9, 0.9}

\resizebox{\linewidth}{!}{
\begin{tabular}{@{} l c cc cc @{\hskip 0.25in} cccc @{\hskip 0.25in} cccc @{}}
\toprule
& & \multicolumn{4}{c}{\textbf{Inference Accuracy} (\%)} & \multicolumn{8}{c}{\textbf{Extraction Metrics} (Precision / Recall) \%} \\
\cmidrule(r{0.25in}){3-6} \cmidrule{7-14}
& & \multicolumn{2}{c}{\textbf{ECHR Dataset}} & \multicolumn{2}{c}{\textbf{Yelp Dataset}} & \multicolumn{4}{c}{\textbf{ECHR Dataset}} & \multicolumn{4}{c}{\textbf{Yelp Dataset}} \\
\cmidrule(lr){3-4} \cmidrule(lr){5-6} \cmidrule(lr){7-10} \cmidrule{11-14}
\textbf{Model} & \textbf{Client} & \cellcolor{secureblue}\textbf{Authorized} $\uparrow$ & \cellcolor{securered}\textbf{Unauthorized} $\downarrow$ & \cellcolor{secureblue}\textbf{Authorized} $\uparrow$ & \cellcolor{securered}\textbf{Unauthorized} $\downarrow$ & \multicolumn{2}{c}{\cellcolor{secureblue}\textbf{Authorized} $\uparrow$} & \multicolumn{2}{c}{\cellcolor{securered}\textbf{Unauthorized} $\downarrow$} & \multicolumn{2}{c}{\cellcolor{secureblue}\textbf{Authorized} $\uparrow$} & \multicolumn{2}{c}{\cellcolor{securered}\textbf{Unauthorized} $\downarrow$} \\
& & (Correct) & (Wrong/No) & (Correct) & (Wrong/No) & Prec & Rec & Prec & Rec & Prec & Rec & Prec & Rec \\
\midrule

\multirow{10}{*}{\rotatebox{90}{\textbf{Qwen~3-1.7B}}} 
& 1  & 22.08 & 4.94 & 26.74 & 4.35 & 1.78 & 17.32 & 10.00 & 4.25 & 2.85 & 6.80 & 0.00 & 0.00 \\
& 2  & 18.07 & 2.63 & 17.05 & 10.39  & 1.57 & 19.15 & 7.71 & 4.96 & 3.77 & 7.50 & 2.05 & 0.97 \\
& 3  & 20.00 & 2.56 & 29.11 & 3.95 & 2.53 & 12.70 & 5.21 & 2.48 & 4.29 & 9.12 & 0.70 & 0.32 \\
& 4  & 20.99 & 5.26 & 24.71 & 8.89 & 2.34 & 14.79 & 3.28 & 1.43 & 4.24 & 9.12 & 1.22 & 0.38 \\
& 5  & 19.78 & 5.68 & 26.83 & 9.72 & 2.38 & 17.21 & 12.14 & 6.82 & 3.36 & 7.78 & 0.85 & 0.38 \\
& 6  & 25.00 & 4.94 & 20.24 & 5.00 & 2.55 & 16.20 & 5.99 & 3.21 & 3.95 & 11.04 & 0.68 & 0.37 \\
& 7  & 22.99 & 2.33 & 21.43 & 4.76 & 2.14 & 18.87 & 6.85 & 4.76 & 3.65 & 7.95 & 1.44 & 0.74 \\
& 8  & 27.85 & 2.60 & 28.21 & 3.90 & 2.44 & 15.56 & 4.32 & 2.74 & 5.66 & 9.67 & 0.99 & 0.36 \\
& 9  & 20.48 & 4.40 & 25.30 & 8.00 & 1.74 & 15.38 & 6.40 & 3.78 & 5.17 & 11.51 & 0.93 & 0.36 \\
& 10 & 23.75 & 1.32 & 17.28 & 6.76 & 2.31 & 12.96 & 5.94 & 4.63 & 2.82 & 8.26 & 0.40 & 0.46 \\
\midrule

\multirow{10}{*}{\rotatebox{90}{\textbf{Gemma~2-2B}}} 
& 1  & 73.26 & 6.02 & 75.00 & 2.41 & 0.00 & 0.00 & 0.00 & 0.00 & 0.27 & 0.97 & 0.29 & 0.97 \\
& 2  & 72.09 & 4.76 & 72.00 & 2.11 & 0.00 & 0.00 & 0.00 & 0.00 & 0.12 & 0.42 & 0.00 & 0.00 \\
& 3  & 57.32 & 4.65 & 71.83 & 3.75 & 0.13 & 0.15 & 0.00 & 0.00 & 0.03 & 0.33 & 0.04 & 0.33 \\
& 4  & 70.89 & 3.75 & 79.47 & 7.95 & 0.00 & 0.00 & 0.00 & 0.00 & 0.03 & 0.35 & 0.00 & 0.00 \\
& 5  & 83.95 & 3.26 & 68.18 & 7.41 & 0.00 & 0.00 & 0.00 & 0.00 & 0.10 & 0.39 & 0.07 & 0.39 \\
& 6  & 61.54 & 4.88 & 75.95 & 3.41 & 0.00 & 0.00 & 0.00 & 0.00 & 0.22 & 0.65 & 0.00 & 0.00 \\
& 7  & 74.16 & 2.25 & 73.61 & 3.66 & 0.44 & 0.18 & 0.00 & 0.00 & 0.17 & 1.52 & 0.05 & 0.38 \\
& 8  & 65.38 & 7.23 & 74.65 & 3.61 & 0.00 & 0.00 & 0.00 & 0.00 & 0.08 & 0.37 & 0.17 & 0.37 \\
& 9  & 64.10 & 6.33 & 71.43 & 3.75 & 0.00 & 0.00 & 0.00 & 0.00 & 0.15 & 0.36 & 0.00 & 0.00 \\
& 10 & 74.03 & 5.63 & 77.27 & 2.44 & 0.00 & 0.00 & 0.00 & 0.00 & 0.00 & 0.00 & 0.10 & 0.46 \\
\midrule

\multirow{10}{*}{\rotatebox{90}{\textbf{Llama~3.2-3B}}} 
& 1  & 44.87 & 7.58 & 50.00 & 5.88 & 3.99 & 12.25 & 8.33 & 0.49 & 3.45 & 8.74 & 0.00 & 0.00 \\
& 2  & 45.33 & 7.58 & 46.03 & 6.67 & 3.07 & 12.48 & 3.00 & 0.51 & 2.11 & 11.25 & 4.55 & 0.42 \\
& 3  & 48.68 & 7.46 & 48.48 & 6.56 & 3.30 & 10.07 & 0.00 & 0.00 & 2.19 & 12.38 & 0.00 & 0.00 \\
& 4  & 43.75 & 10.34 & 60.56 & 2.94 & 3.66 & 11.92 & 1.82 & 0.16 & 3.41 & 8.77 & 8.82 & 1.05 \\
& 5  & 51.35 & 2.60 & 51.61 & 12.77 & 2.70 & 16.56 & 9.52 & 0.97 & 2.13 & 10.12 & 0.00 & 0.00 \\
& 6  & 46.58 & 7.81 & 49.25 & 8.20 & 3.90 & 13.28 & 4.76 & 0.44 & 3.14 & 11.69 & 8.70 & 0.65 \\
& 7  & 36.00 & 1.49 & 57.38 & 14.29 & 3.48 & 15.34 & 0.00 & 0.00 & 1.70 & 12.88 & 1.92 & 0.38 \\
& 8  & 34.38 & 1.89 & 60.61 & 9.09 & 3.77 & 12.82 & 4.44 & 0.68 & 3.04 & 10.04 & 8.33 & 1.12 \\
& 9  & 44.59 & 6.06 & 43.40 & 4.76 & 3.81 & 11.19 & 4.55 & 0.42 & 3.79 & 17.63 & 3.85 & 0.36 \\
& 10 & 43.48 & 4.29 & 48.21 & 6.67 & 3.15 & 11.88 & 4.17 & 0.31 & 1.92 & 9.17 & 3.12 & 0.46 \\
\bottomrule
\end{tabular}
}\vspace{-5mm}
\end{table*}

\noindent\textbf{\textsc{SecureGate} vs. Baselines.}
We compare \textsc{SecureGate} against standard FL baselines— FedAvg, FedAdagrad, FedAdam, FedYogi, and FedAvgM (\sref{sec:FL_LLMs})—under inference attack accuracy with 10 clients. We integrate \textsc{SecureGate}'s dual-adapter architecture and token-controlled gating into each baseline and evaluate the resulting \emph{SecureGate-enhanced} methods (Table~\ref{tab:fl_full_comparison_no_rotate_main}). 

With valid tokens, \textsc{SecureGate} achieves an average inference accuracy of 25.20\%, outperforming the strongest baseline (FedAvgM at 20.12\%). Without authorization, it maintains low leakage (4.20\%), comparable to restrictive baselines such as FedYogi (3.73\%), while delivering substantially higher authorized utility. These results demonstrate that \textsc{SecureGate} consistently improves the privacy--utility trade-off across both classical and personalized FL settings.

\noindent\textbf{Evaluating \textsc{SecureGate}'s PII-Aware Preservation.}
We evaluate \textsc{SecureGate} under extraction and inference attacks on ECHR and Yelp using Qwen~3-1.7B, Gemma~2-2B, and Llama~3.2-3B, comparing correct versus incorrect or missing tokens. As shown in Table~\ref{tab:combined_results}, inference attacks with {\em valid tokens}, Gemma~2-2B achieves the highest inference accuracy (57.32--83.95\% on ECHR; 68.18--79.47\% on Yelp), followed by Llama~3.2-3B and Qwen3-1.7B. This highlights that while Gemma resists explicit memorization, it remains susceptible to probabilistic, ranking-based inference. In {\em unauthorized settings}, inference leakage remains low across all models (1.32--10.34\% on ECHR; 2.11--14.29\% on Yelp). Notably, authorization removal causes sharp performance drops—for instance, Llama~3.2-3B (Client~5, ECHR) falls from 51.35\% to 2.60\% ($19.75\times$ reduction), and Gemma~2-2B (Client~10, Yelp) from 77.27\% to 2.44\% ($31.66\times$ reduction).

For extraction attacks with {\em valid tokens}, all models exhibit low precision. Qwen~3-1.7B attains precision of 1.57--2.55\% (ECHR) and 2.82--5.66\% (Yelp), while Llama~3.2-3B ranges from 2.70--3.99\% (ECHR) and 1.70--3.79\% (Yelp).  Corresponding recall values remain modest, peaking at 19.15\% (Qwen~3-1.7B, ECHR). Gemma~2-2B shows the lowest susceptibility to verbatim extraction, consistent with its memorization-mitigation training~\cite{peng2024data}, though it remains vulnerable to probabilistic, ranking-based inference.

When {\em tokens are incorrect or absent}, all models demonstrate strong resistance to extraction, with recall dropping near zero. For instance, Llama~3.2-3B recall decreases to at most 1.12\% on Yelp, representing up to a $17.07\times$ reduction relative to authorized access. Similar trends hold for Qwen~3-1.7B, proving that the gating module reliably suppresses sensitive disclosures without authorization.

Overall, these results show that \textsc{SecureGate} enforces access-aware privacy: sensitive information is revealed only with valid authorization, while incorrect or missing tokens reliably trigger a secure state. Candidate pool scaling ($c=50,100,500$) is analyzed in Appendix~\ref{subsec:candidate_scaling}.

\begin{figure}[!t]
\centering
    \includegraphics[width=\linewidth]{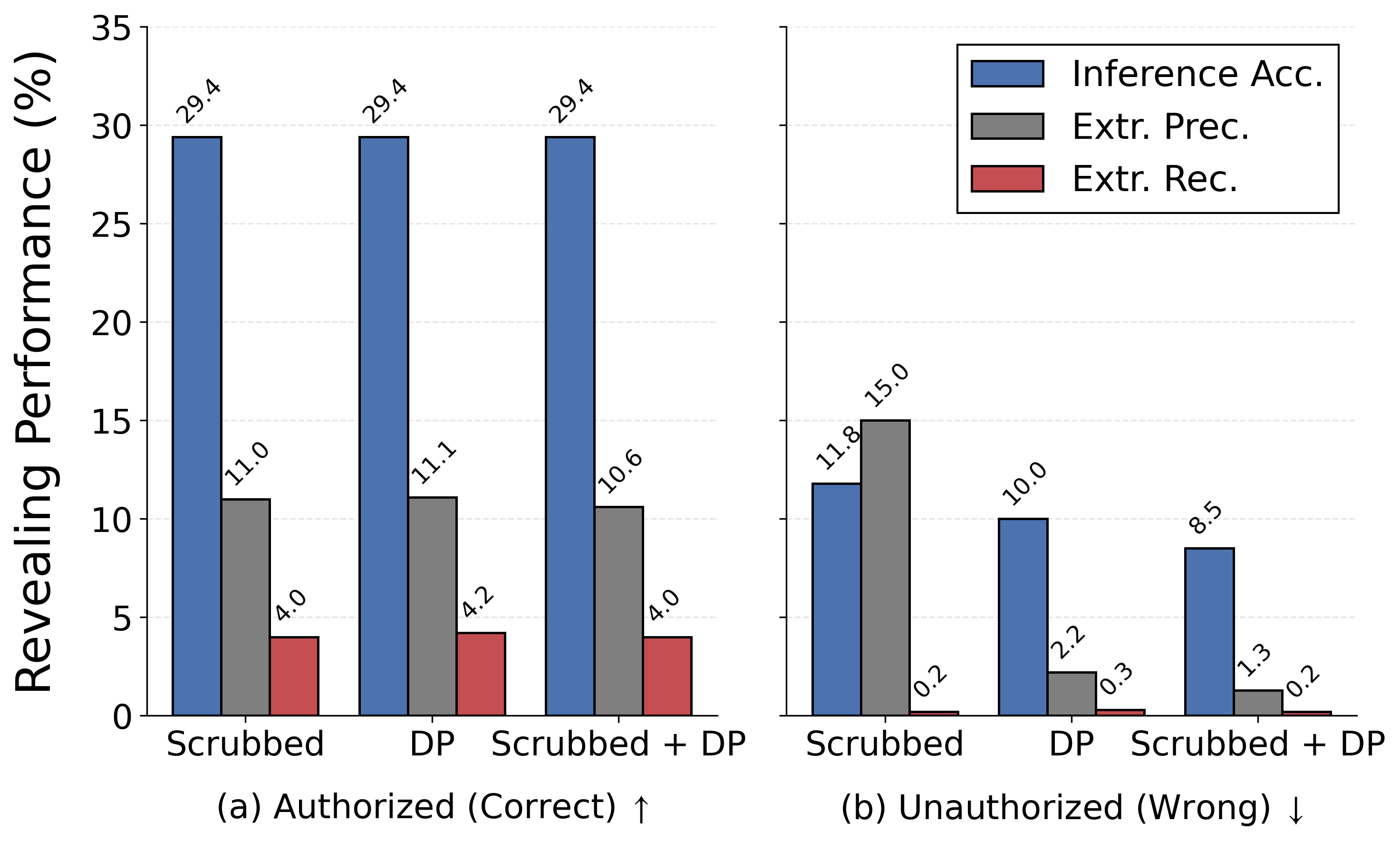}
\caption{Gating module performance in routing queries to the correct adapter under inference and extraction attacks across various defenses on Llama-1B.}
\label{fig:securegate_defence_analysis}\vspace{-5mm}
\end{figure}

\vspace{1mm}
\noindent\textbf{Evaluating the Effectiveness of \textsc{SecureGate}'s Gating Model Across Defenses.}
We evaluate \textsc{SecureGate} per client under multiple defense strategies (data scrubbing, DP, and scrubbing+DP).
As shown in \fref{fig:securegate_defence_analysis}a, with a {\em valid authorization token}, the gating module achieves \(100\%\) routing accuracy to the revealing adapter, yielding a consistent \(29.41\%\) revealing performance across all attacks, since defenses are not activated.

For {\em unauthorized queries}, the gating module reliably routes requests to the secure adapter. As illustrated in \fref{fig:securegate_defence_analysis}b, the hybrid \textsc{Scrubbed+DP} defense offers the strongest protection, reducing extraction precision to \(1.30\%\) (\(11.54\times\) lower than scrubbing alone at \(15.00\%\)) and inference accuracy to \(8.51\%\) (\(3.46\times\) lower than the authorized baseline). The perfect routing accuracy confirms that any residual leakage stems from the defense mechanisms themselves rather than failures in token-based gating. Overall, \textsc{SecureGate} effectively integrates layered defenses to achieve strong privacy guarantees while preserving authorized utility.

\noindent\textbf{Multi-Adapter Evaluation: Authorization, Isolation, and Privacy Floors}\label{sec:multi_adapter} To assess how well \textsc{SecureGate} manages dataset-specific and role-based permissions, we tested a configuration featuring several ``revealing'' adapters alongside a single ``secure'' adapter. In this Llama-1B setup, each client hosts four specialized heads: one adapter fine-tuned on ECHR, one on Yelp, a joint adapter trained on the union of ECHR and Yelp (``E + Y''), and a secure adapter trained on a scrubbed, combined corpus where PII has been masked. Each adapter is bound to a distinct keyed token, and the router is trained to map every key deterministically to its corresponding adapter, thereby implementing per-key access control over personalization paths.

Table \ref{tab:llama1b_singleclient_inference_attack_multi_adapter} summarizes these outcomes, showing that the gating module achieves 100\% routing reliability by successfully directing all requests to the authorized adapter path without collision. Crucially, the results reveal a significant \textit{Authorization Gap}; for the ECHR dataset, an attacker with valid credentials achieves an inference accuracy of 20.00\%, which drops to a \textit{Privacy Floor} of 2.44\% when no valid Key is provided. This represents an $8.20\times$ reduction in PII leakage for unauthorized users. There is also clear \textit{cross-dataset isolation}: using the Yelp key on ECHR candidates reaches just 4.94\% accuracy, and using incorrect or missing keys on Yelp tops out at 6.76\%, indicating that miskeyed requests expose only a small fraction of the sensitive information that is available under the correct key. Finally, the joint ``E + Y'' adapter attains 13.41\% on ECHR—lower than the 20.00\% of the dedicated ECHR adapter—which suggests that mixing heterogeneous sources during training can dilute dataset-specific memorization and slightly dampen leakage.

\begin{table*}[!t]
\centering
\caption{Inference attack accuracy on a single client using Llama-1B. The diagonal (light blue) demonstrates perfect adapter routing under valid authorization keys.}\vspace{-2mm}
\label{tab:llama1b_singleclient_inference_attack_multi_adapter}
\footnotesize
\setlength{\tabcolsep}{15pt} 
\renewcommand{\arraystretch}{1.3}
\definecolor{secureblue}{rgb}{0.9, 0.95, 1.0}

\resizebox{\linewidth}{!}{%
\begin{tabular}{ll c cccc}
\toprule
 & & & \multicolumn{4}{c}{\textbf{Adapter Selected} (Routing Result)} \\
\cmidrule(lr){4-7}
\textbf{Dataset} & \textbf{Auth Key Used} & \textbf{Inference Accuracy} $\uparrow$ & \textbf{E + Y} & \textbf{ECHR} & \textbf{Yelp} & \textbf{Masked} \\
\midrule
\multirow{4}{*}{\textbf{ECHR}} & ECHR \& Yelp Key & 13.41 & \cellcolor{secureblue}100.00 & 0.00 & 0.00 & 0.00 \\
                               & ECHR Key         & 20.00 & 0.00 & \cellcolor{secureblue}100.00 & 0.00 & 0.00 \\
                               & Yelp Key          & 4.94  & 0.00 & 0.00 & \cellcolor{secureblue}100.00 & 0.00 \\
                               & Wrong/No Key      & 2.44  & 0.00 & 0.00 & 0.00 & \cellcolor{secureblue}100.00 \\
\midrule
\multirow{4}{*}{\textbf{Yelp}} & ECHR \& Yelp Key & 12.20 & \cellcolor{secureblue}100.00 & 0.00 & 0.00 & 0.00 \\
                               & ECHR Key         & 5.19  & 0.00 & \cellcolor{secureblue}100.00 & 0.00 & 0.00 \\
                               & Yelp Key          & 13.92 & 0.00 & 0.00 & \cellcolor{secureblue}100.00 & 0.00 \\
                               & Wrong/No Key      & 6.76  & 0.00 & 0.00 & 0.00 & \cellcolor{secureblue}100.00 \\
\bottomrule
\end{tabular}%
}\vspace{-3mm}
\end{table*}

\begin{figure}[!t]
\centering
\includegraphics[width=1\linewidth]
{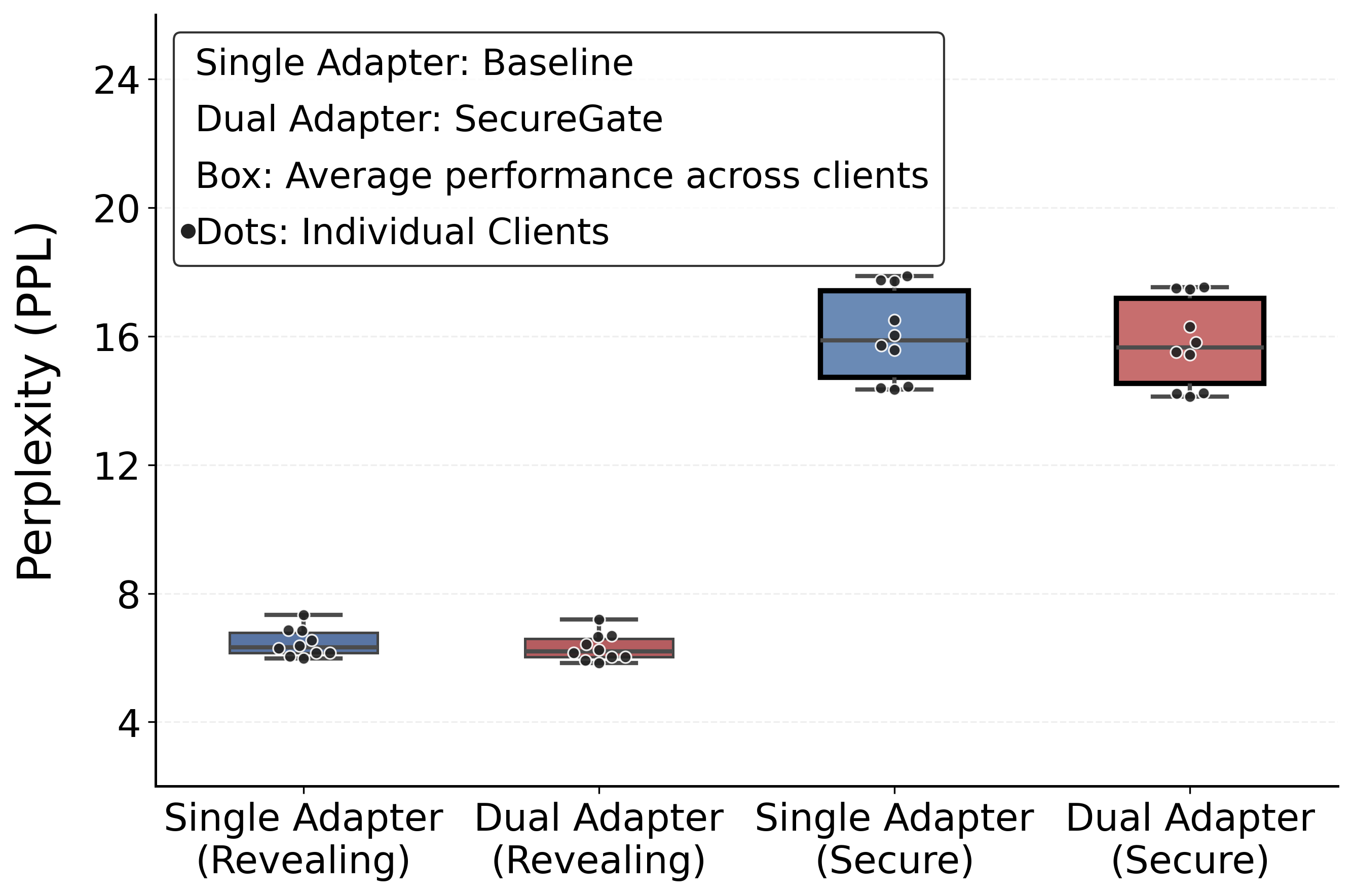}
\caption{PPL across ten clients for Llama-1B, showing that \textsc{SecureGate} matches the utility of standalone secure and revealing adapter baselines.}
    \label{fig:perplexity_analysis}
    \vspace{-5mm}
\end{figure}
\vspace{1mm}
\noindent\textbf{Perplexity Analysis and Utility Preservation.}
\fref{fig:perplexity_analysis} compares \textsc{SecureGate}'s dual-adapter design with single-adapter baselines (secure or revealing) on Llama-1B using the ECHR dataset. With a valid token, \textsc{SecureGate} achieves an average Perplexity (PPL) of 6.32, matching the revealing-adapter baseline (6.46). Under unauthorized access (missing or incorrect token), it reliably switches to the secure adapter, yielding a PPL of 15.89, close to the secure-adapter baseline of 16.03. Detailed results in Table~\ref{tab:ppl_direct_comparison} (Appendix~\ref{subsec:detailed_perplexity_results}) confirm that the gating mechanism preserves {\em personalized model utility} while enforcing privacy constraints.

\begin{figure}[!t]
\centering
\includegraphics[width=1\linewidth]{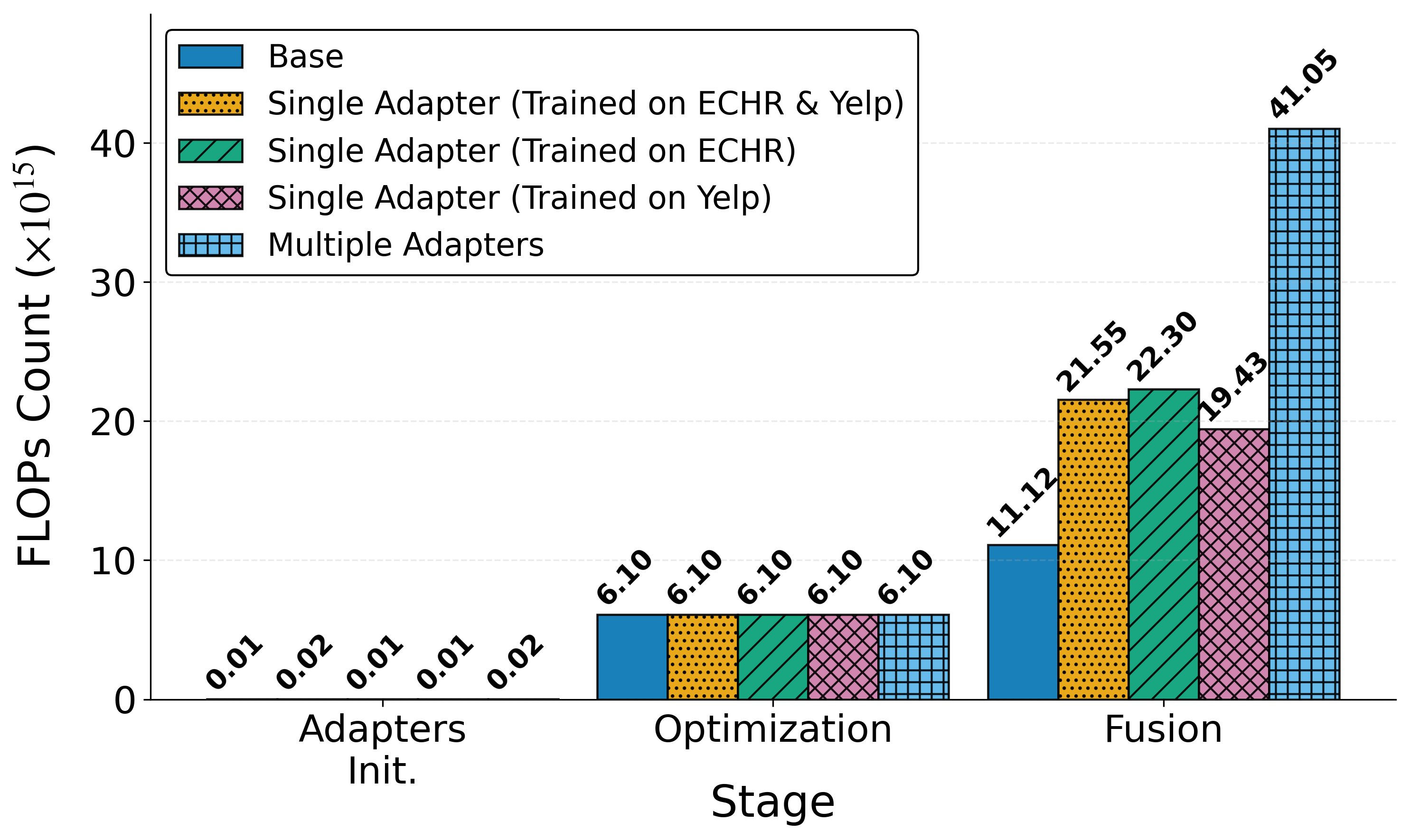}
\caption{Client-side computational cost across the three Federated LLM fine-tuning stages (initialization, optimization,  fusion), showing that multi-adapter configurations increase overhead only during the fusion stage.}
\label{fig:trainable_flops}
\vspace{-6mm}
\end{figure}

\noindent\textbf{Computational Cost Analysis.}
\fref{fig:trainable_flops} reports the phase-wise FLOPs across four configurations. The optimization phase remains constant at $6.10\times10^{15}$ FLOPs, while the fusion phase scales with the number of revealing adapters, increasing from $11.12\times10^{15}$ to $41.05\times10^{15}$ FLOPs ($3.69\times$). This shows that computational overhead is dominated by local adapter fusion rather than server-side training or communication. A detailed configuration and communication cost is provided in Appendix~\ref{subsec:paramter_overhead}.

\noindent\textbf{Gating Overhead Analysis}
\label{subsec:gating_overhead} We evaluate the computational overhead introduced by the gating mechanism in \textsc{SecureGate} compared to a single-adapter baseline. As shown in Table~\ref{tab:llama_echr_timing}, the baseline performs only the adapter pass, whereas \textsc{SecureGate} introduces an additional gating pass prior to adapter execution.

The gating pass incurs a fixed cost of 197.14 ms, while the adapter pass remains unchanged across both baseline and \textsc{SecureGate} configurations. This results in a modest overhead of approximately $1.13\times$ per inference, consistent across both secure and revealing settings.

Notably, the gating computation is lightweight compared to the adapter pass, contributing only a small fraction of the total latency. These results demonstrate that \textsc{SecureGate} achieves controlled and predictable overhead, making it practical for deployment in real-world systems where both privacy guarantees and efficiency are critical.

\begin{table}[!t]
\centering
\caption{Gating overhead analysis: Llama 1B gated adapters vs single-adapter baseline (ECHR dataset)}\vspace{-2mm}
\label{tab:llama_echr_timing}
\footnotesize
\setlength{\tabcolsep}{4pt}
\renewcommand{\arraystretch}{1.3}

\definecolor{gatinggreen}{rgb}{0.9, 1.0, 0.9}
\definecolor{adapterorange}{rgb}{1.0, 0.95, 0.85}

\resizebox{\columnwidth}{!}{%
\begin{tabular}{l cc p{0.1cm} cc}
\toprule
& \multicolumn{2}{c}{\textbf{Baseline}} 
&& \multicolumn{2}{c}{\textbf{\textsc{SecureGate} (Ours)}} \\
\cmidrule(lr){2-3} \cmidrule(lr){5-6}
& \textbf{Secure} & \textbf{Revealing} 
&& \textbf{Secure} & \textbf{Revealing} \\
\midrule

\textbf{Time (ms)} (\(\downarrow\)) 
& \multicolumn{2}{c}{} 
&& \multicolumn{2}{c}{} \\

\textbf{Gating Pass} 
& -- & -- 
&& \cellcolor{gatinggreen}197.14 
& \cellcolor{gatinggreen}197.14 \\

\textbf{Adapter Pass} 
& \cellcolor{adapterorange}1501.71 
& \cellcolor{adapterorange}1463.55 
&& \cellcolor{adapterorange}1501.71 
& \cellcolor{adapterorange}1463.55 \\

\midrule

\textbf{Gating Overhead} (\(\times\)) 
& 1$\times$ & 1$\times$ 
&& \textbf{1.13$\times$} & \textbf{1.13$\times$} \\

\bottomrule
\end{tabular}
}\vspace{-4mm}
\end{table}

\vspace{-1mm}
\section{Conclusion}
 \vspace{-1mm}


We propose \textsc{SecureGate}, a privacy-aware federated fine-tuning framework for LLMs that ensures strong privacy, high utility, and efficiency. It combines a dual-adapter architecture with token-controlled gating to enable fine-grained personalization while protecting sensitive PII from inference and extraction attacks. Extensive evaluations show that \textsc{SecureGate} achieves strong privacy–utility trade-offs across heterogeneous settings with minimal overhead, offering a scalable and practical approach to secure federated LLMs.
\clearpage
\section{Limitations}
Despite the effectiveness of \textsc{SecureGate}, several practical limitations remain. The additional ``gating pass'' executed prior to adapter inference increases the token prefill latency, which may affect deployment in latency-sensitive, real-time applications. Maintaining multiple active LoRA adapters (secure, revealing, and global) also incurs additional memory overhead compared to single-adapter personalization, potentially limiting adoption on resource-constrained edge devices. Although \textsc{SecureGate} achieves \(100\%\) routing reliability in our experiments, its security ultimately depends on the secrecy and uniqueness of key-token representations, motivating future work on stronger cryptographic embedding mechanisms to further harden access control at scale. Finally, our evaluation focuses on decoder-only architectures (e.g., Llama, Qwen, Gemma), and extending the proposed dual-adapter design to substantially larger models or alternative architectures remains an open research direction.



\bibliography{custom}

\newpage
\clearpage
\appendix

\section*{Appendices}
\label{appendix:more_results}

\appendix
\section{Detailed Experimental Settings}
\label{sec:experimental_details}

\subsection{Reproduction Hyperparameters}
To ensure reproducibility, all LoRA adapters were configured with a scaling factor of \(4 \times r\) and a dropout rate of 0.1. Weight decay was set to \(\gamma = 0.001\), and global aggregation applied a momentum of \(m = 0.5\). Personalized adapter fusion used a regularization factor \(\psi = 0.01\). All other experimental parameters follow \cite{lukas2023analyzing}.

\subsection{Evaluation Metrics}
\label{subsec:metric_definitions}
We describe the evaluation metrics used to assess privacy, utility, and computational efficiency as:

\begin{itemize}[leftmargin=*]
    \item \textbf{Precision:} Fraction of generated PII that matches the training data, indicating actual leakage.
    \item \textbf{Recall:} Fraction of training PII successfully reproduced, reflecting potential leakage risk.
    \item \textbf{FLOPs:} Total floating-point operations, independent of hardware; lower values indicate higher computational efficiency.
    \item \textbf{Perplexity (PPL):} Measures model uncertainty in predicting text; lower values indicate better predictive performance:
    \[\Resize{7cm}{
    \text{PPL} = \exp\Bigg(-\frac{1}{H} \sum_{i=1}^{H} \log P(\kappa_i \mid \kappa_1, \dots, \kappa_{i-1})\Bigg),}
    \]
    where \(H\) is the total number of tokens and \(P(\kappa_i \mid \kappa_1, \dots, \kappa_{i-1})\) is the predicted probability of token \(\kappa_i\).
\end{itemize}

\subsection{Gating Module Architecture}
\label{subsec:gating_module}
The design of our \textsc{SecureGate}'s gating module comprises:
\begin{itemize}[leftmargin=*]
    \item A linear layer with GLU activation, followed by layer normalization and dropout.
    \item A second linear layer with GELU activation, layer normalization, and dropout for feature refinement.
    \item A final linear layer producing logits over available adapters.
\end{itemize}
GLU dynamically controls feature flow for routing, while GELU introduces smooth non-linear transformations to stabilize training. Together, they enable flexible and efficient adapter selection.

\subsection{Computing Environment}
\label{sec:hardware_appendix}
All experiments ran on Ubuntu 24.04.2 LTS with an Intel Core i9-14900K CPU @ 3.2GHz and an NVIDIA RTX PRO 6000 Blackwell GPU (96GB VRAM). This configuration supports concurrent loading of multiple LoRA adapters in \textsc{SecureGate}.

\subsection{Implementation Details}
\label{subsec:implementation}
The framework is implemented using OpenFedLLM~\cite{ye2024openfedllm}. Named Entity Recognition (NER) employs the Flair library~\cite{akbik2019flair}, using hyperparameters consistent with \cite{lukas2023analyzing} to ensure comparability and reproducibility.


\begin{table}[!t]
\centering
\caption{Inference attack accuracy (\%) for Llama-1B across non-IID clients (Clients 1--5: ECHR, Clients 6--10: Yelp) under correct-token and wrong/no-token settings.}

\label{tab:non_iid_inference_attack}
\footnotesize
\renewcommand{\arraystretch}{1.2}
\definecolor{secureblue}{rgb}{0.9, 0.95, 1.0}
\definecolor{securered}{rgb}{1.0, 0.9, 0.9}
\setlength{\tabcolsep}{5pt}
\resizebox{\columnwidth}{!}{%
\begin{tabular}{l l cc}
\toprule
\textbf{Client \#} & \textbf{Training Dataset} & \cellcolor{secureblue}\textbf{Correct Token} & \cellcolor{securered}\textbf{Wrong/No Token} \\
\midrule
1  & ECHR & 36.00 & 5.33 \\
2  & ECHR & 33.33 & 4.48 \\
3  & ECHR & 31.33 & 6.10 \\
4  & ECHR & 22.37 & 10.67 \\
5  & ECHR & 31.65 & 4.94 \\
\midrule
6  & Yelp & 17.91 & 1.59 \\
7  & Yelp & 32.00 & 6.94 \\
8  & Yelp & 24.68 & 7.79 \\
9  & Yelp & 18.31 & 3.23 \\
10 & Yelp & 12.86 & 6.06 \\
\bottomrule
\end{tabular}
}
\end{table}

\section{In-Depth Evaluation of \textsc{SecureGate}}

\subsection{Inference Performance under Non-IID Client Distributions} \label{subsec:non_iid}

Table~\ref{tab:non_iid_inference_attack} reports inference attack accuracy (\%) for Llama-1B across {\em ten non-IID} clients (organizations), where clients~1--5 are trained on ECHR and clients~6--10 on Yelp, explicitly modeling a heterogeneous federated LLMs setting.

Under \emph{authorized access} (correct token), ECHR clients exhibit consistently higher  revealing percentage (inference success), reaching up to 36.00\% (Client~1) and remaining above 30\% for several clients (Clients~2, 3, and~5). In contrast, most Yelp clients achieve lower accuracy, with some below 25\% (e.g., 17.91\% for Client~6 and 12.86\% for Client~10). This suggests that models trained on structured legal text are more vulnerable to inference attacks than those trained on informal review data. Nevertheless, notable attack success on some Yelp clients (e.g., 32.00\% for Client~7 and 24.68\% for Client~8) confirms that data heterogeneity does not eliminate the threat but instead induces client-specific vulnerability profiles.

Under \emph{unauthorized access} (wrong or missing token), inference accuracy drops sharply across all clients, typically to the 1--11\% range (e.g., 5.33\% for Client~1, 10.67\% for Client~4, and 1.59\% for Client~6), demonstrating the effectiveness of token-controlled gating in suppressing attacks by routing queries to the secure adapter. A small number of outliers (e.g., Client~4 at 10.67\% and Client~8 at 7.79\%) still exhibit residual leakage, indicating that the privacy strength ultimately depends on the underlying defense used in the secure adapter. These results underscore the need for stronger or client-adaptive defenses to ensure uniform privacy guarantees under highly non-IID data distributions.

\begin{figure}[!t]
  \centering
  \includegraphics[width=1\linewidth]{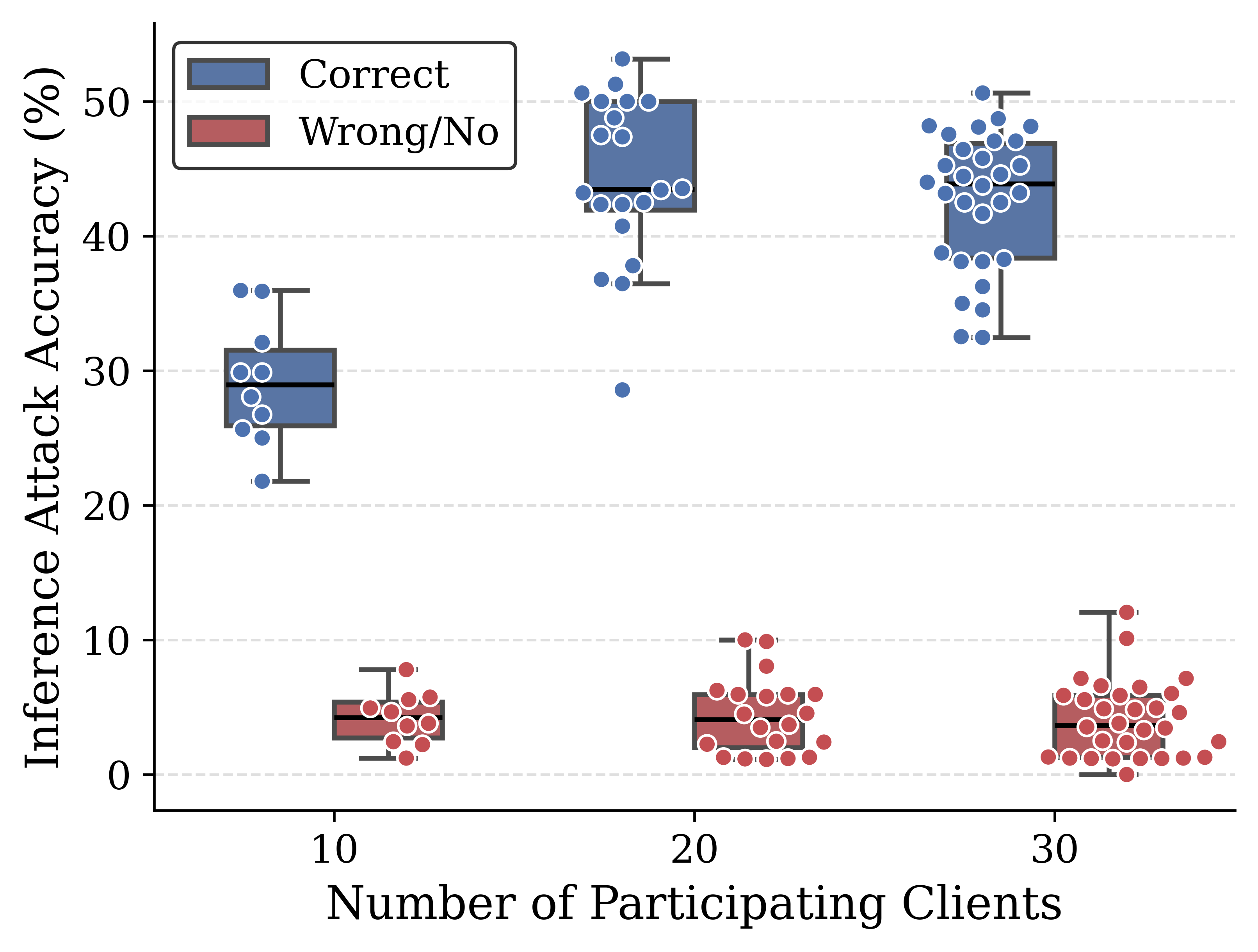}
  \caption{Scalability analysis of inference attack accuracy for Llama~3.2-1B across 10, 20, and 30 clients. The gating mechanism successfully decouples authorized utility from unauthorized leakage, maintaining a stable privacy boundary ($\approx$ 4.2\%) as the network scales.}
  \label{fig:echr_llama32_1b_different_clients}
\end{figure}

\subsection{Evaluating Scalability Across Client Distributions}

To assess the scalability and robustness of our \textsc{SecureGate}, \fref{fig:echr_llama32_1b_different_clients} reports inference attack accuracy for the Llama~3.2--1B model under increasing numbers of federated clients. Under authorized access (correct token), the mean attack success rate rises from 29.09\% ($\sigma=4.62$) with 10 clients to 44.32\% ($\sigma=6.23$) with 20 clients, reflecting improved model utility as additional client data contributes to fine-tuning. 

In contrast, under unauthorized access, leakage remains consistently low and largely invariant to network scale. Mean attack success rates are 4.20\% ($\sigma=1.95$) for 10 clients, 4.36\% ($\sigma=2.83$) for 20 clients, and 4.11\% ($\sigma=2.86$) for 30 clients. This stability demonstrates that \textsc{SecureGate} maintains strong privacy guarantees as the federated network scales, ensuring that increased client participation does not introduce additional privacy risk.

\begin{figure}[!t]
    \centering
    \begin{subfigure}[b]{1\linewidth}
        \centering
        \includegraphics[width=\linewidth]{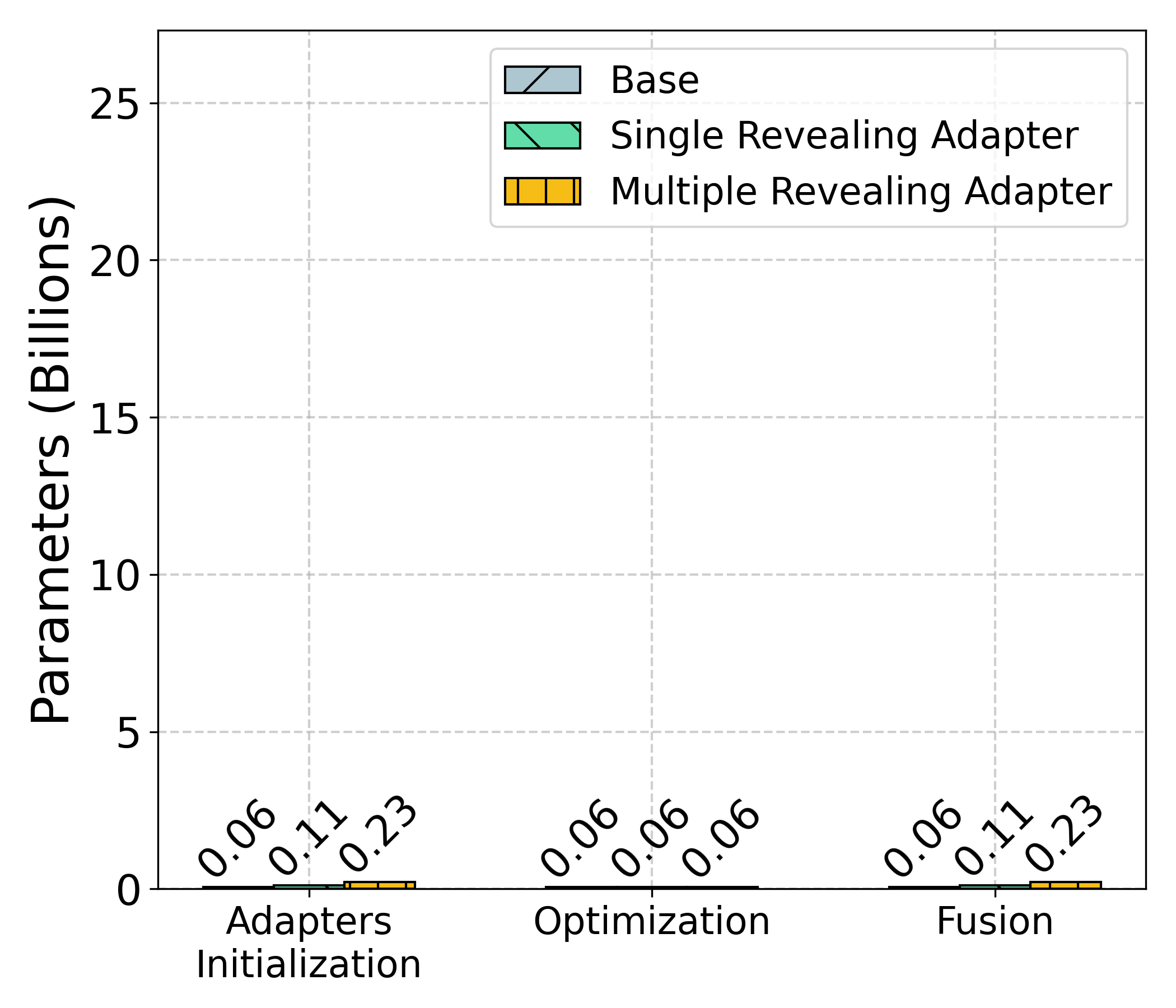}
        \caption{Trainable Parameters} 
        \label{fig:trainable_params}
    \end{subfigure}
    \\[1ex] 
    \begin{subfigure}[b]{1\linewidth}
        \centering
        \includegraphics[width=\linewidth]{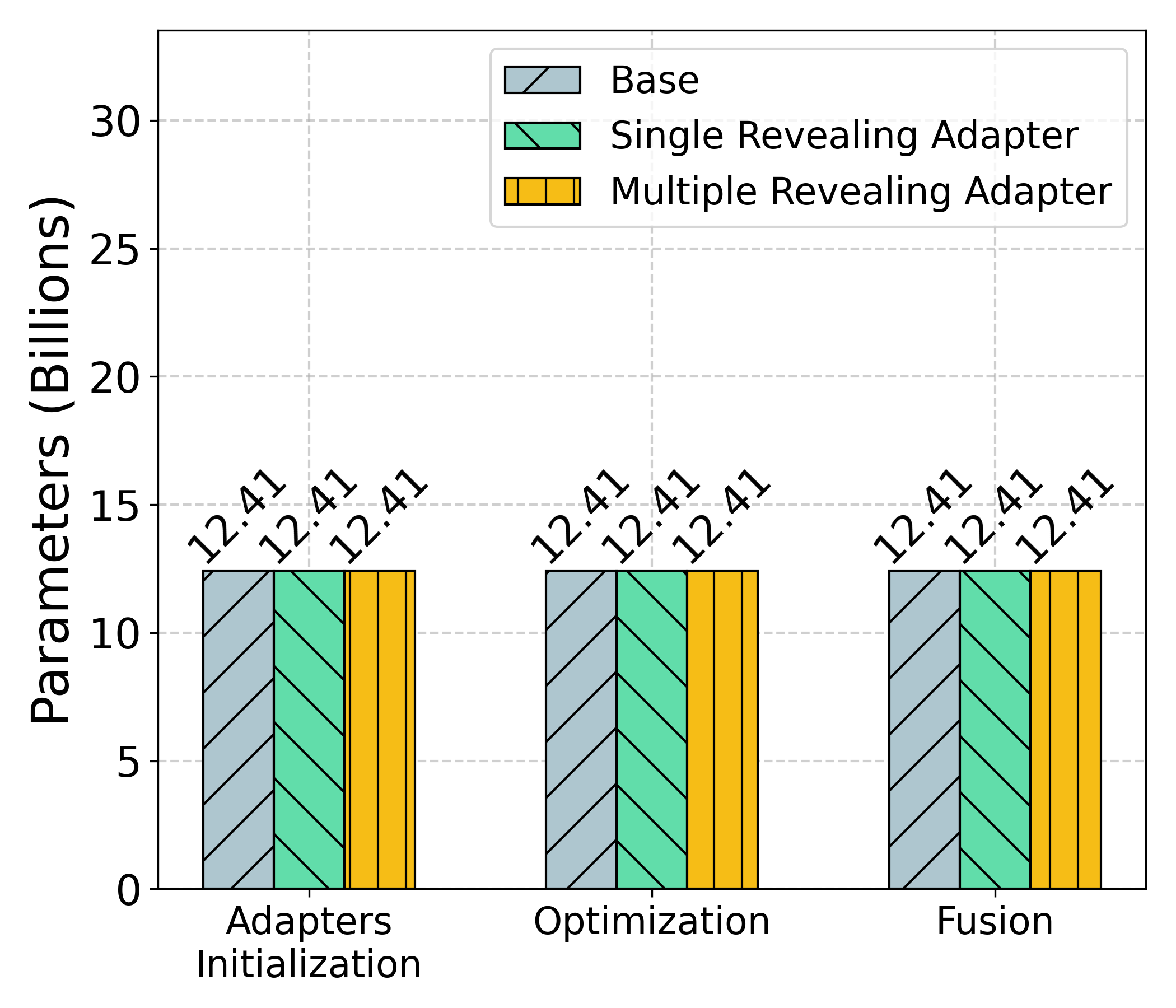}
        \caption{Non-Trainable Parameters} 
        \label{fig:frozen_params}
    \end{subfigure}
    \caption{Comparative analysis of trainable vs. non-trainable parameter counts (billions) across framework stages. The analysis demonstrates the minimal footprint of authorization-specific adapters relative to the frozen base model. Values are rounded to two decimals.}
    \label{fig:combined_params}
\end{figure}

In summary, these results showcase two core strengths of \textsc{SecureGate}. Authorized users with correct tokens retain full access to personalized knowledge, while any unauthorized or miskeyed requests are automatically isolated within a secure adapter that minimizes PII exposure. This per-request gating provides a robust, auditable access control framework that protects sensitive data without degrading legitimate model utility.

\subsection{Evaluating \textsc{SecureGate} Communication Overhead}
\label{subsec:paramter_overhead}
To further evaluate the overhead of the proposed authorization mechanism, we analyze the parameter distribution across the operational phases of \textsc{SecureGate} using the LlaMA~3.2-1B model. This analysis builds directly upon the previously described authorization experiment (see~\sref{sec:multi_adapter}) conducted on the ECHR and Yelp datasets, comparing base performance against configurations with single or multiple revealing adapters. As shown in Figure \ref{fig:combined_params}, the frozen parameters remain constant at 12.41 billion regardless of setup, while trainable parameters rise from 0.06 billion to 0.23 billion during the initialization and fusion phases when scaling to multiple revealing adapters, a 3.83$\times$ increase in local capacity. Crucially, the optimization phase remains unaffected by these configurations, maintaining a minimal trainable footprint of 0.06 billion. By confining the training of multiple personalized adapters to the client side, the mechanism minimizes communication overhead and prevents interference with global training.



\begin{table}[!t]
\centering
\caption{Performance comparison across PII classes for correct and wrong/no token cases, along with adapter selection ratios, on inference attack accuracy (\%) for Llama-3B on the Yelp dataset.}
\label{tab:llama3b_yelp_inference_pii_correct_wrong}
\footnotesize 
\setlength{\tabcolsep}{1pt}
\renewcommand{\arraystretch}{1.3}
\definecolor{secureblue}{rgb}{0.9, 0.95, 1.0}
\definecolor{securered}{rgb}{1.0, 0.9, 0.9}

\resizebox{\linewidth}{!}{%
\begin{tabular}{ l cc p{0.1cm} cc}
\toprule
& \multicolumn{2}{c}{\textbf{\cellcolor{secureblue} Correct Token} (\(\uparrow\))} 
&& \multicolumn{2}{c}{\cellcolor{securered}\textbf{Wrong/No Token} (\(\downarrow\))} \\ 
\cmidrule{2-3} \cmidrule{5-6}
\textbf{PII Class} & \textbf{Acc. (\%)} & \textbf{Reveal Adapter} && \textbf{Acc. (\%)} & \textbf{Secure Adapter} \\ 
\midrule
PERSON  & 51.72 & \cellcolor{secureblue!60}100.0\% && 5.36  & \cellcolor{securered!60}100.0\% \\
LOC     & 77.14 & \cellcolor{secureblue!60}100.0\% && 24.24  & \cellcolor{securered!60}100.0\% \\
ORG     & 84.91 & \cellcolor{secureblue!60}100.0\% && 10.71 & \cellcolor{securered!60}100.0\% \\
PRODUCT & 85.71 & \cellcolor{secureblue!60}100.0\% && 6.90  & \cellcolor{securered!60}100.0\% \\
\midrule
\textbf{Mean} & \textbf{74.87} & \textbf{100.0\%} && \textbf{11.80} & \textbf{100.0\%} \\
\bottomrule
\end{tabular}%
}
\end{table}

\subsection{Inference Attack Performance by PII Class}
\label{subsec:inferencet_pii_class}
Table~\ref{tab:llama3b_yelp_inference_pii_correct_wrong} reports inference attack accuracy of Llama-3B on the Yelp dataset across different PII classes under Correct and Wrong/No Tokens. When provided the correct token, attack accuracy is high for all PII categories (PERSON, LOC, ORG, PRODUCT), with a mean accuracy of \(74.87\%\), indicating that the model reliably reveals targeted PII when properly authorized. Under Wrong/No Tokens, accuracy drops to the \(5.36\%\)–\(24.24\%\) range with a mean of \(11.80\%\), showing that the defended adapter substantially suppresses unauthorized PII extraction, though location entities (LOC) remain relatively more vulnerable than other PII types. Notably, our routing mechanism achieves a \(100.0\%\) selection rate for the intended adapter (Reveal or Secure) across all PII classes. This perfect adapter selection confirms that the residual inference accuracy observed in categories such as LOC (\(24.24\%\)) is not due to routing failures, but rather the inherent difficulty of masking location-based context even within the secure state.

\subsection{Effect of Candidate Pool Size on Inference Accuracy}
\label{subsec:candidate_scaling}

Table~\ref{tab:llama_echr_date} illustrates the relationship between the number of candidates (\(c\)) and inference attack accuracy for the ECHR Date class. As expected, increasing the candidate pool size from \(c=50\) to \(c=500\) results in a consistent reduction in accuracy for both token conditions. While correct authorization provides higher accuracy across all scales, the unauthorized accuracy (Wrong/No Token) drops significantly as the search space expands, reaching its lowest point (\(4.49\%\)) at \(c=500\). This trend highlights that the defense becomes increasingly robust against brute-force or high-entropy inference attacks as the candidate pool grows.

\begin{table}[!t]
\centering
\caption{Inference attack accuracy (\%) for Llama~3.2-3B on ECHR (Date class) across different candidate sizes (\(c\)) for a single client.}
\label{tab:llama_echr_date}
\footnotesize
\setlength{\tabcolsep}{5pt}
\renewcommand{\arraystretch}{1.3}
\definecolor{secureblue}{rgb}{0.9, 0.95, 1.0}
\definecolor{securered}{rgb}{1.0, 0.9, 0.9}
\resizebox{\columnwidth}{!}{%
\begin{tabular}{@{} l ccc @{}}
\toprule
& \multicolumn{3}{c}{\textbf{Candidate Pool Size ($c$)}} \\
\cmidrule(l){2-4}
\textbf{Token Condition} & \textbf{\(c=50\)} & \textbf{\(c=100\)} & \textbf{\(c=500\)} \\
\midrule
\cellcolor{secureblue}Correct Token (\(\uparrow\))  & 15.85 & 10.10 & 7.45 \\
\cellcolor{securered}Wrong/No Token (\(\downarrow\)) & 10.84 & 8.33  & 4.49 \\
\bottomrule
\end{tabular}
}
\end{table}

\begin{table*}[!t]
\centering
\caption{Inference attack accuracy (\%) for Llama-1B on the ECHR dataset under different LoRA ranks (\(r\)) and numbers of clients.}
\label{tab:llama1b_echr_inference_lora_clients}
\footnotesize
\setlength{\tabcolsep}{12pt}
\renewcommand{\arraystretch}{1.2}
\definecolor{secureblue}{rgb}{0.9, 0.95, 1.0}
\definecolor{securered}{rgb}{1.0, 0.9, 0.9}

\resizebox{\linewidth}{!}{
\begin{tabular}{lccccp{0.1cm}cccc}
\toprule
\textbf{Client} &
\multicolumn{4}{c}{\cellcolor{secureblue}\textbf{Correct Token}} &&
\multicolumn{4}{c}{\cellcolor{securered}\textbf{Wrong/No Token}} \\
\cmidrule{2-5} \cmidrule{7-10}
\# & \(r=4\) & \(r=8\) & \(r=12\) & \(r=16\) && \(r=4\) & \(r=8\) & \(r=12\) & \(r=16\) \\
\midrule

1  & 25.64 & 27.16 & \cellcolor{secureblue!60}\textbf{40.48} & 35.96 && 3.66 & \cellcolor{securered!60}4.71 & 1.52 & 3.61 \\
2  & 18.42 & 24.64 & 20.25 & \cellcolor{secureblue!60}\textbf{25.64} && 1.64 & 0.00 & 4.84 & \cellcolor{securered!60}5.56 \\
3  & 15.28 & 20.97 & 22.50 & \cellcolor{secureblue!60}\textbf{26.74} && 1.35 & 2.47 & 4.41 & \cellcolor{securered!60}4.65 \\
4  & 17.33 & 22.03 & \cellcolor{secureblue!60}\textbf{26.32} & 21.79 && \cellcolor{securered!60}11.39 & 6.49 & 6.35 & 4.94 \\
5  & 27.27 & 25.76 & 25.00 & \cellcolor{secureblue!60}\textbf{29.87} && 5.00 & 2.50 & \cellcolor{securered!60}6.94 & 2.22 \\
6  & 17.11 & 18.31 & 17.86 & \cellcolor{secureblue!60}\textbf{25.00} && 2.94 & \cellcolor{securered!60}9.09 & 6.56 & 3.80 \\
7  & 17.11 & 20.83 & 21.43 & \cellcolor{secureblue!60}\textbf{28.05} && 3.53 & 3.66 & 0.00 & \cellcolor{securered!60}5.75 \\
8  & 20.29 & 22.73 & 27.16 & \cellcolor{secureblue!60}\textbf{29.87} && 1.59 & 2.70 & 1.72 & \cellcolor{securered!60}7.79 \\
9  & 22.22 & \cellcolor{secureblue!60}\textbf{36.23} & 30.38 & 32.10 && 1.37 & 0.00 & \cellcolor{securered!60}3.03 & 1.22 \\
10 & 25.97 & 33.33 & 22.35 & \cellcolor{secureblue!60}\textbf{35.90} && \cellcolor{securered!60}2.50 & 2.47 & 1.43 & 2.44 \\
\bottomrule
\end{tabular}
}
\end{table*}

\begin{table}[!t]
\centering
\caption{Client-wise Perplexity (PPL) comparison between \textsc{SecureGate} and standalone revealing or secure adapter baselines. Values are rounded to two decimals.}
\label{tab:ppl_direct_comparison}
\scriptsize
\setlength{\tabcolsep}{1pt}
 \renewcommand{\arraystretch}{1}
\definecolor{highlight}{gray}{0.93}
\definecolor{securegateblue}{rgb}{0.9, 0.95, 1.0}
\resizebox{\linewidth}{!}{
\begin{tabular}{lccp{0.1cm}cc}
\toprule
& \multicolumn{2}{c}{\textbf{Standalone Baselines} } && \multicolumn{2}{c}{\textbf{\textsc{SecureGate} (Ours)} } \\
\cmidrule{2-3} \cmidrule{5-6}
\textbf{Client} & \textbf{Revealing} & \textbf{Secure} & &\cellcolor{securegateblue}\textbf{Correct Token} & \textbf{Wrong/No Token} \\
\midrule
1  & 6.86 & 16.02 && \cellcolor{securegateblue}\textbf{6.69} & 15.81 \\
2  & 6.03 & 17.71 && \cellcolor{securegateblue}\textbf{5.92} & 17.49 \\
3  & 5.97 & 15.57 && \cellcolor{securegateblue}\textbf{5.83} & 15.43 \\
4  & 6.55 & 15.71 && \cellcolor{securegateblue}\textbf{6.42} & 15.50 \\
5  & 6.29 & 14.34 && \cellcolor{securegateblue}\textbf{6.16} & 14.12 \\
6  & 6.85 & 17.87 && \cellcolor{securegateblue}\textbf{6.66} & 17.53 \\
7  & 6.16 & 14.44 && \cellcolor{securegateblue}\textbf{6.03} & 14.22 \\
8  & 7.33 & 17.74 && \cellcolor{securegateblue}\textbf{7.20} & 17.47 \\
9  & 6.14 & 14.38 && \cellcolor{securegateblue}\textbf{6.02} & 14.23 \\
10 & 6.37 & 16.50 && \cellcolor{securegateblue}\textbf{6.24} & 16.29 \\
\midrule
\rowcolor{highlight}
\textbf{Average} & \textbf{6.46} & \textbf{16.03} && \cellcolor{securegateblue}\textbf{6.32} & \textbf{15.89} \\
\bottomrule
\end{tabular}
}
\end{table}

\section{Ablation Study of \textsc{SecureGate} Components}

\subsection{Evaluating LoRA Rank Impact} 
Table~\ref{tab:llama1b_echr_inference_lora_clients} evaluates the impact of LoRA ranks ($r=4, 8, 12, 16$) on the robustness of \textsc{SecureGate} against inference attacks using the Llama-1B model. When the correct token is utilized, inference accuracy generally increases with higher LoRA ranks, peaking at $r=16$ for most clients, such as Client 10 reaching 35.90\%. Notably, Client 1 achieves the highest overall authorized disclosure of 40.48\% at $r=12$. In unauthorized scenarios using wrong/no tokens, leakage remains consistently low, typically staying below 10\% across all ranks.

The defense remains robust even as model capacity increases; for instance, at $r=16$, Client 8's accuracy drops from 29.87\% (Correct) to 7.79\% (Wrong), a reduction of 22.08 percentage points and approximately $3.83\times$ lower success. While some fluctuations occur—such as Client 4 reaching 11.39\% at $r=4$—the leakage at higher ranks like $r=16$ often remains more stable, as seen with Client 10's 2.44\% leakage. This disparity highlights that for some users, higher ranks actually reduce the risk of unintended exposure compared to $r=4$, where Client 4 sees a $2.31\times$ higher leakage risk. This indicates that while higher ranks can improve authorized utility, the gating module effectively maintains a secure state regardless of the LoRA rank; however, it is important to highlight that effective LoRA privacy tuning necessitates careful rank selection, tailored to each client's specific requirements.

\subsection{Language Modeling Performance (Perplexity)}
\label{subsec:detailed_perplexity_results}
Table \ref{tab:ppl_direct_comparison} presents the full perplexity (PPL) metrics for all 10 clients using the Llama-1B model on the ECHR dataset. These results compare our \textsc{SecureGate} framework against standalone revealing and secure adapter baselines. The data confirms that \textsc{SecureGate} maintains linguistic utility similar to the standalone adapters while successfully enforcing token-based privacy boundaries.

\begin{figure}[t]
    \centering
    \includegraphics[width=\linewidth]{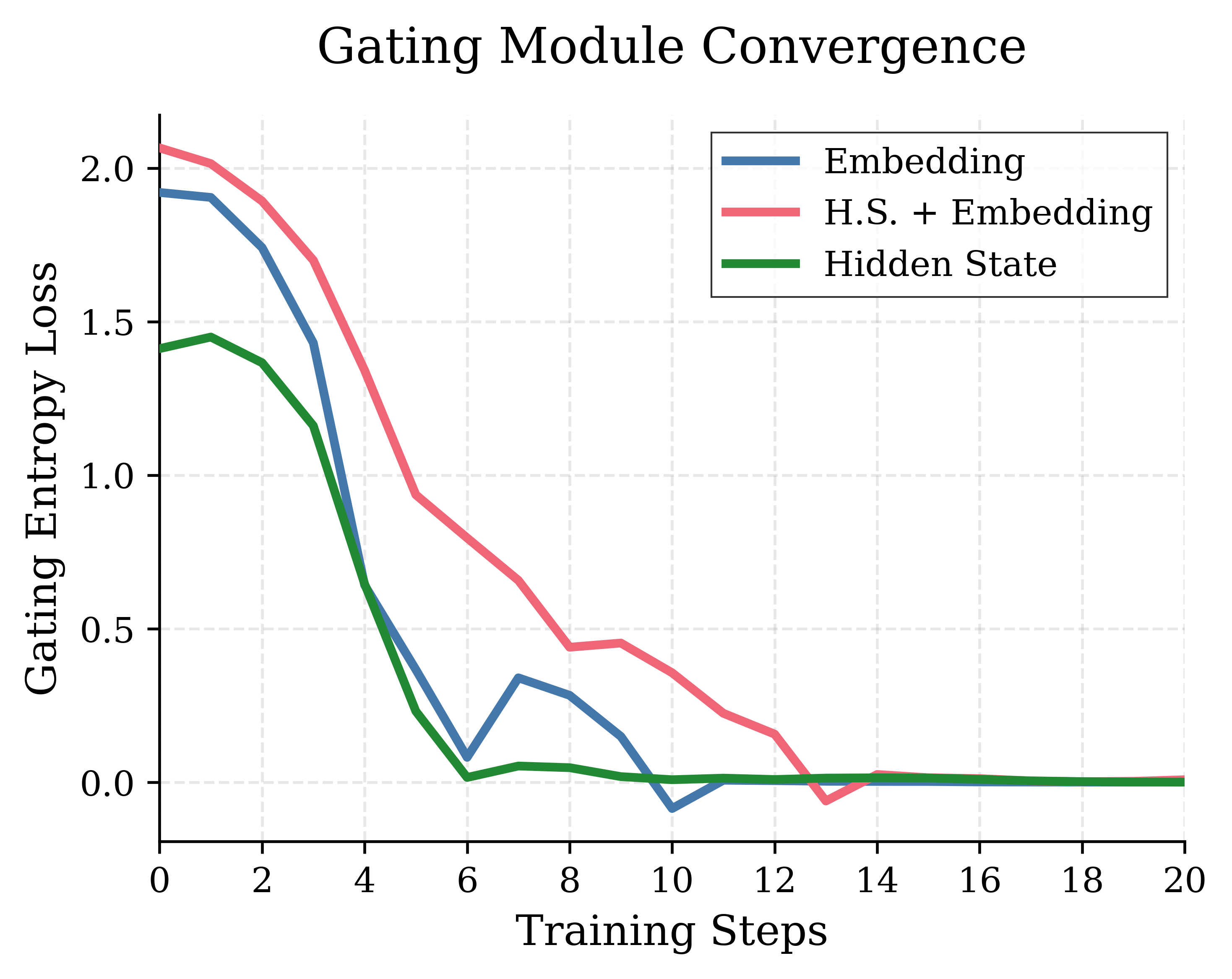}
    \caption{Training loss convergence of the \textsc{SecureGate} gating module across different input features. The rapid convergence within 20 steps highlights the module's lightweight computational requirements.}
    \label{fig:gating_feature_ablation}
\end{figure}

\subsection{Gating Feature Analysis}
\label{subsec:gating_feature_ablation}

We evaluate three candidate feature sources for the gating module: (i) raw token embeddings, (ii) final-layer hidden states ($h_{\text{key}}$), and (iii) their concatenation. Although all configurations converge within 20 training steps, using only the final-layer hidden states consistently yields the most reliable routing signal. This choice results in lower gating entropy, more accurate adapter selection, and improved training stability, while avoiding the increased dimensionality and overfitting risks introduced by feature concatenation. 

Based on these observations, we adopt the final-layer hidden state as the default gating input in the \textsc{SecureGate} architecture. \fref{fig:gating_feature_ablation} illustrates the corresponding gating entropy loss curves, demonstrating both rapid convergence and a steady reduction in routing uncertainty during training.

\section{Security Analysis of Keyed Tokenization and Routing}
\label{subsec:key-security}

We consider an adversary with full access to \emph{public} model artifacts (base weights, public tokenizer, public logs, and prompts) but without privileged access to organization-specific keys, local tokenizers, or key-management systems. The adversary may (i) inspect or modify public prompts, (ii) attempt to infer or brute-force keys from public artifacts, and (iii) observe adapter routing via black-box queries.

\noindent\textbf{Security Properties.}
Under this threat model, \textsc{SecureGate} provides the following guarantees:

\begin{enumerate}[leftmargin=*]
    \item \textbf{Key confidentiality.} Keys exist only in local tokenizers and are absent from public vocabularies, preventing exposure or enumeration by external tools.
    
    \item \textbf{Atomic key representation.} Each key is a single vocabulary entry and never split into subtokens, preventing partial leakage via subtoken analysis.
    
    \item \textbf{Deterministic, non-revealing embeddings.} Key embeddings \(e_{\text{key}} \in \mathbb{R}^d\) are generated from a keyed hash and pseudorandom generator, making inversion computationally infeasible.
    
    \item \textbf{Low-cost, deterministic routing.} A compact gating MLP maps key embeddings to adapter indices deterministically, supporting reproducible, auditable routing.
    
    \item \textbf{Efficient key rotation and revocation.} Key updates only require replacing the keyed embedding and updating the gating module, without retraining the base model or adapters.
    
    \item \textbf{Resistance to tokenization-based attacks.} Public tokenizers never contain key tokens, mitigating prompt-scraping, token-guessing, and vocabulary-based attacks.
    
    \item \textbf{Operational robustness.} Keys, local tokenizers, and gating checkpoints are managed under least-privilege control, with versioning and secure backups to enable safe rotation and rollback.
\end{enumerate}

\noindent\textbf{Discussion.} 
Items (1)–(2) ensure keys remain atomic and secret within the organization, preventing leakage via public tooling. Item (3) provides reproducible embeddings that are opaque to outsiders. Items (4)–(5) guarantee deterministic, auditable routing while supporting rapid key rotation. Finally, items (6)–(7) reduce the attack surface against token-based, prompt-manipulation, and operational attacks, ensuring secure adapter selection without exposing secrets.

\subsection{Deterministic, Non-Revealing Embeddings}
Special tokens representing authorization keys are added to the tokenizer vocabulary as $T = \{t_1, \dots, t_N\}$, where each $t_k \in \Sigma^L$ is a unique string of length $L$ over alphanumeric $\Sigma$. In \textsc{SecureGate}, we use two tokenizers: a \emph{public tokenizer} that never contains key tokens and can be shared across organizations, and a \emph{local private tokenizer} maintained by each organization that stores the protected keys after hashing. The local tokens are first generated as plaintext strings and then hashed before being added into the private tokenizer. The key space entropy is
\begin{equation}
H(K) = L \log_2 |\Sigma|,
\end{equation}
which results in a key space of $|\Sigma|^L$ and provides exponential growth in entropy.

The base seed is derived from the full set of key tokens:
\begin{equation}
S_{\mathrm{base}} = \left\lfloor \frac{1}{N} \sum_{k=1}^N \mathrm{hash}(t_k) \right\rfloor,
\end{equation}
where $\mathrm{hash}(t_k)$ extracts a fixed-length value from the MD5 hexadecimal digest of string $t_k$ and converts it into an integer. This $S_{\mathrm{base}}$ functions as a collective key derived from all tokens.

For each token $t_k \in T$, we compute
\begin{equation}
S_{t_k} = \mathrm{hash}(t_k).
\end{equation}
\begin{equation}
S_{\mathrm{comb}} = S_{\mathrm{base}} + S_{t_k}.
\end{equation}

The key embedding $\mathbf{e}_{\mathrm{key}}(t_k) \in \mathbb{R}^d$ is generated as
\begin{equation}
\mathbf{e}_{\mathrm{key}}(t_k) \sim \mathcal{N}(0, \sigma^2 I_d) \mid S_{\mathrm{comb}}.
\end{equation}

In our deployment, we use $L = 20$ alphanumeric characters with $|\Sigma| = 62$ (uppercase A–Z, lowercase a–z, and digits 0–9), yielding $H(K) \approx 119$ bits. The brute-force success probability with $Q$ queries is
\begin{equation}
P_{\text{success}} = \min\left(1, \frac{Q}{|\Sigma|^L}\right).
\end{equation}
For $Q = 10^9$, this yields approximately $2^{-89}$, rendering brute-force attacks infeasible under realistic query budgets.

This transformation ensures deterministic key embeddings while maintaining practical non-invertibility. Specifically, recovering authorization strings from $\mathbf{e}_{\mathrm{key}}$ is infeasible due to information loss introduced by hashing and the stochastic embedding process.

\section{Synthetic Data Generation and Augmentations for Gating Training}
\label{subsec:synthetic_routing_data}

The routing dataset $\mathcal{D}_{\text{gate}}$ is constructed to systematically cover nominal, malformed, and adversarial input conditions. It includes: (i) valid keyed prompts; (ii) malformed keys; (iii) empty-key positions; and (iv) no-key baselines. For the purpose of these examples, identifiers such as \texttt{ALPHA}, \texttt{BETA}, and \texttt{GAMMA} serve as human-readable placeholders for the high-entropy, secret keys actually utilized by the system. When the gating module encounters an invalid or unauthorized key, the system is designed to deterministically fall back to a designated secure adapter. The following examples characterize the diverse input types within $\mathcal{D}_{\text{gate}}$ used to train the gating module for robust routing and fallback detection:

\begin{algorithm}[!t]
\caption{\textsc{SecureGate} Training and Personalized Fusion}
\label{alg:securegate_training}
\KwIn{
    Raw client datasets $\{\mathcal{D}_n^{\text{raw}}\}_{n=1}^N$; Frozen base model $M$; \\
    Local secure tokenizer $\mathcal{T}_{sec}$
}
\KwOut{
    Personalized adapters $\{\Delta \boldsymbol{w}_{p,n}^{(\text{sec})}, \Delta \boldsymbol{w}_{p,n}^{(\text{rev})}\}$; Trained gating MLP $G$
}

\textbf{(1) Local Initialization (Sec. 4.1):}
\begin{enumerate}[label=\arabic*., noitemsep, topsep=0pt]
    \item Generate masked datasets $\mathcal{D}_n^{\text{mask}}$ via NER and scrubbing.
    \item Initialize revealing and secure adapters $\Delta \boldsymbol{w}_{n}^{(\text{rev})}, \Delta \boldsymbol{w}_{n}^{(\text{sec})}$ following \eqref{eq:lora_initalize}.
\end{enumerate}

\textbf{(2) Federated Fine-Tuning (Sec. 4.2):}
\begin{enumerate}[label=\arabic*., noitemsep, topsep=0pt]
    \item \textbf{Inner Loop:} Minimize local objective on $\mathcal{D}_n^{\text{mask}}$ to produce $\Delta \boldsymbol{w}_n^{t+1}$.
    \item \textbf{Outer Loop:} Aggregate secure updates $\Delta \bar{\boldsymbol{w}}^{t+1}$ using \eqref{eq:outer_loop}.
    \item Update momentum $\Delta \boldsymbol{v}^{t+1}$ via \eqref{eq:momentem_based}.
    \item Compute global model $\Delta\boldsymbol{w}^{t+1}$ using \eqref{eq:global_update_using_momentum}.
\end{enumerate}

\textbf{(3) Personalized Fusion (Sec. 4.3):}
\begin{enumerate}[label=\arabic*., noitemsep, topsep=0pt]
    \item Fuse secure personalized adapter $\Delta\boldsymbol{w}^{(\text{sec})}_{p,n}$ using \eqref{eq:secure_fusion}.
    \item Fuse revealing personalized adapter $\Delta\boldsymbol{w}^{(\text{rev})}_{p,n}$ using \eqref{eq:reveal_fusion}.
    \item Optimize fusion coefficients $(\alpha, \beta)$ via \eqref{eq:fusion_obj}.
\end{enumerate}

\textbf{(4) Gating Module Training (Sec. 4.4):}
\begin{enumerate}[label=\arabic*., noitemsep, topsep=0pt]
    \item Train $G$ on synthetic key patterns using the cross-entropy loss
$\mathcal{L}_{\text{gating}} = -\sum_{i} y_i \log p_i$.

\end{enumerate}
\end{algorithm}

\noindent{\bf Illustrative Data Examples:}
The following samples demonstrate the diverse input types within $\mathcal{D}_{\text{gate}}$ used to harden the routing mechanism:

\begin{description}[leftmargin=*]
    \item[Valid Key Tokens:] Representations of authorized requests using well-formed keys.
    \begin{itemize}[leftmargin=*]
        \item \texttt{[SPECIAL\_TOKEN=ALPHA]} Example input for module processing
        \item \texttt{[SPECIAL\_TOKEN=BETA]} A prompt designed to assess token routing
        \item \texttt{[SPECIAL\_TOKEN=GAMMA]} Request requiring special token handling
    \end{itemize}

    \item[Malformed/Incorrect Keys:] Samples designed to test the router's resistance to corruption.
    \begin{itemize}[leftmargin=*]
        \item \texttt{[SPECIAL\_TOKEN=ALPH]} Example input for module processing
        \item \texttt{[SPECIAL\_TOKEN=BETA\_CORRUPT]} Scenario of ambiguous adapter routing
    \end{itemize}

    \item[Empty/No Key Baselines:] Cases that must trigger the default fallback logic.
    \begin{itemize}[leftmargin=*]
        \item \texttt{[SPECIAL\_TOKEN=]} Empty string placeholder
        \item \texttt{[SPECIAL\_TOKEN= ]} An instruction that triggers fallback adapter
        \item A prompt designed to assess token routing
    \end{itemize}
\end{description}

\noindent By training across these distinct key states, the gating network learns to enforce a deterministic mapping for valid authorization tokens while identifying corrupted or missing keys as candidates for the secure fallback adapter.

\section{Algorithms}
\label{subsec:algorithms}

Algorithm \ref{alg:securegate_training} provides the complete training workflow for the \textsc{SecureGate} framework, which is executed in four primary stages: (1) {\em local initialization} involving data masking through NER and scrubbing, (2) {\em federated fine-tuning} utilizing inner-loop local updates and outer-loop momentum-based aggregation, (3) {\em the personalized fusion} of secure and revealing adapters, and (4) {\em the final optimization} of the gating module. 

Algorithm \ref{alg:securegate_inference} details the two-pass inference process. The first pass focuses on gating and authorization by extracting hidden states and enforcing security policies through a confidence threshold \(\tau\), while the second pass performs clean generation by stripping key tokens to prevent output contamination.

\begin{algorithm}[!t]
\caption{\textsc{SecureGate} Two-Pass Inference Gating}
\label{alg:securegate_inference}
\KwIn{
    User request $X_{raw}$; Personalized adapters $\{\Delta \boldsymbol{w}_{p,n}^{(\text{sec})}, \Delta \boldsymbol{w}_{p,n}^{(\text{rev})}\}$; \\
    Gating MLP $G$; Confidence threshold $\tau$
}
\KwOut{
    Authorized or sanitized model response $Y$
}

\textbf{Pass 1: Gating \& Authorization}
\begin{enumerate}[label=\arabic*., noitemsep, topsep=0pt]
    \item Forward request through $M$ and extract hidden state $h_{\text{key}}$ via \eqref{eq:hidden_extraction}.
    \item Compute selection probabilities $p_i$ using the softmax
$p_i = \frac{\exp(z_i)}{\sum_{j\in\mathcal{I}} \exp(z_j)}$.

    \item Select adapter index $a^* = \arg\max p_i$.
    \item \textbf{Enforce Policy:} If $p_{a^*} \le \tau$, default to secure adapter $\Delta \boldsymbol{w}_{p,n}^{(\text{sec})}$.
\end{enumerate}

\textbf{Pass 2: Clean Generation}
\begin{enumerate}[label=\arabic*., noitemsep, topsep=0pt]
    \item Strip key token from prompt to prevent output contamination.
    \item Generate final response $Y$ using selected adapter $\Delta \boldsymbol{w}_{p,n}^{(a^*)}$.
\end{enumerate}

\Return $Y$
\end{algorithm}

\end{document}